\documentclass[prd,preprint,eqsecnum,nofootinbib,amsmath,amssymb,
tightenlines]{revtex4}

\usepackage{graphicx}
\usepackage{bm}
\usepackage{amsmath}
\usepackage{amssymb}   
\usepackage{latexsym}  
\usepackage{color}

\newcommand{\Tr}{\mbox{${\rm Tr}$}}

\newcommand{\comment}[1]{}

\newcommand{\lsim}{\mbox{\raisebox{-0.6ex}{$\stackrel{<}{\sim}$}}\:}

\begin{document}

\title{Ultraviolet avalanche in anisotropic non-Abelian plasmas}
\author{Adrian Dumitru$^a$, Yasushi Nara$^a$ and Michael Strickland$^b$}
\affiliation{$^a$ Institut f\"ur Theoretische Physik,
Johann Wolfgang Goethe Universit\"at,
Max von Laue Str.\ 1,
D-60438 Frankfurt am Main, Germany\\
$^b$ Frankfurt Institute for Advanced Study, 
Johann Wolfgang Goethe Universit\"at,
Max von Laue Str.\ 1,
D-60438 Frankfurt am Main, Germany}
\begin{abstract}
We present solutions of coupled particle-field evolution in classical
$U(1)$ and $SU(2)$ gauge theories in real time on three-dimensional
lattices. Our simulations are performed in a regime of extreme
anisotropy of the momentum distribution of hard particles where
back-reaction is important. We find qualitatively different behavior
for the two theories when the field strength is high enough that
non-Abelian self-interactions matter for $SU(2)$. It appears that the
energy drained by a Weibel-like plasma instability from the particles
does not build up exponentially in soft transverse magnetic fields but
instead returns, isotropically, to the hard scale via a rapid
avalanche into the ultraviolet.
\end{abstract}
\date{\today}
\maketitle

\section{Introduction}

One of the most important open questions emerging from the
Relativistic Heavy-Ion Collider (RHIC) program is how quickly, and by
what process, the stress-energy tensor of the high-density QCD matter
produced in the central rapidity region becomes isotropic. This is
important, for example, to determine the maximal temperature achieved
in such collisions, which will also be pursued in the near future at
even higher energies at the CERN Large Hadron Collider (LHC).
One of the chief obstacles to thermalization in ultrarelativistic
heavy-ion collisions is the rapid longitudinal expansion of the
central rapidity region. If the matter expands too quickly then there
will not be sufficient time for its constituents to interact and
thermalize.  In the absence of interactions, the longitudinal
expansion causes the system to become much colder in the longitudinal
direction than in the transverse directions, corresponding to $\langle
p_L^2 \rangle \ll \langle p_T^2\rangle$ in the local rest frame.  One
can then ask how long it would take for interactions to restore
isotropy in momentum space.

High-energy heavy-ion collisions release a large amount of partons
from the wavefunctions of the colliding nuclei. Partons with very
large transverse momenta originate from high-$Q^2$ hard interactions,
while partons with transverse momenta below the so-called saturation
momentum $Q_s$ (given by the square root of the color charge density
per unit area in the incoming nuclei) are much more abundant and are
better viewed as a classical non-Abelian
field~\cite{Mueller:2002kw,Iancu:2003xm,McLerran:2005kk}.  At high
energies, $Q_s\gg\Lambda_{\rm QCD}$ sets a semi-hard scale which
allows for a weak-coupling treatment of the early-stage
dynamics. However, due to the large occupation number of the soft
modes below $Q_s$ (the classical color field has strength
$F_{\mu\nu}\sim1/g$), the problem is nevertheless non-perturbative.

The evolution following the initial impact was among the questions
which the so-called ``bottom-up scenario''~\cite{Baier:2000sb}
attempted to answer. For the first time, it addressed the dynamics of
soft modes (``fields'') with momenta much below $Q_s$ coupled to the
hard modes (``particles'') with momenta on the order of $Q_s$ and
above. However, it has emerged recently that one of the assumptions
made in this model is not correct.  The debate centers around the very
first stage of the bottom-up scenario in which it was assumed that (a)
collisions between the high-momentum (or hard) modes were the driving
force behind isotropization and that (b) the low-momentum (or soft)
fields act only to screen the electric interaction. In
doing so, the bottom-up scenario implicitly assumed that the
underlying soft gauge modes behaved the same in an anisotropic plasma
as in an isotropic one.  However, it turns out that in anisotropic
plasmas, within the so-called ``hard loop approximation'' (see below),
the most important collective mode corresponds to an instability to
transverse magnetic field fluctuations
\cite{Mrowczynski:1993qm,Mrowczynski:1994xv,Mrowczynski:1996vh}.
Recent works have shown that the presence of these instabilities is
generic for distributions which possess a momentum-space anisotropy
\cite{Romatschke:2003ms,Arnold:2003rq,Romatschke:2004jh} and have
obtained the full hard-loop (HL) action in the presence of an
anisotropy \cite{Mrowczynski:2004kv}.  Another important development
has been the demonstration that such instabilities exist in numerical 
solutions to classical Yang-Mills fields in an expanding geometry 
\cite{Romatschke:2005ag,Romatschke:2005pm,Romatschke:2006nk}.

In the last year there have been significant advances in the
understanding of non-Abelian soft-field dynamics in anisotropic
plasmas within the HL
framework~\cite{Arnold:2005vb,Rebhan:2005re,Romatschke:2006wg}.  The
HL framework is equivalent to the collisionless Vlasov theory of
eikonalized hard particles, i.e.\ the particle trajectories are
assumed to be unaffected by the induced background field. It is
strictly applicable only when there is a very large scale separation
between the soft and hard momentum scales.\footnote{Of course, in the
hard-loop approach very small angle deflections, $\theta \sim g$, are
included self-consistently.}  Even with these simplifying assumptions,
HL dynamics for self-interacting gauge fields is non-trivial due to
the presence of non-linear interactions which can act to regulate
unstable growth.  These non-linear interactions become important when
the vector potential amplitude is on the order of $A_{\rm non-Abelian}
\sim p_{\rm s}/g \sim \sqrt{f_h} p_h$, where $p_h$ is the characteristic
momentum of the hard particles, $f_h$ is the angle-averaged hard
occupancy, and $p_s \sim g \sqrt{f_h} p_h$ is the characteristic soft
momentum of the fields.\footnote{We will not attempt here to use the
specific fields obtained from the Color Glass Condensate wave
functions~\cite{Krasnitz:1998ns,Krasnitz:1999wc,Krasnitz:2000gz,Krasnitz:2001qu,Krasnitz:2002mn,Krasnitz:2003jw,Krasnitz:2002ng,Lappi:2003bi,Lappi:2006fp,Fries:2006pv}
as initial conditions. Hence, we label the hard particle momentum
scale more generically as $p_h$ rather than $Q_s$.} In QED there is no
such complication and the fields grow exponentially until $A_{\rm
Abelian} \sim p_h/g$ at which point the hard particles undergo
large-angle deflections by the soft background field invalidating the
assumptions underpinning the hard-loop effective action.

Recent numerical studies of HL gauge dynamics for SU(2) gauge theory
indicate that for {\em moderate} anisotropies the gauge field dynamics
changes from exponential field growth indicative of a conventional
Abelian plasma instability to linear growth when the vector potential
amplitude reaches the non-Abelian scale, $A_{\rm non-Abelian} \sim
p_{\rm h}$ \cite{Arnold:2005vb,Rebhan:2005re}. This linear growth
regime is characterized by a cascade of the energy pumped into the
soft modes by the instability to higher-momentum plasmon-like modes
\cite{Arnold:2005ef,Arnold:2005qs}. These results indicate that there
is a fundamental difference between Abelian and non-Abelian plasma
instabilites.  However, even with this new understanding the HL
framework relies on the existence of a very large separation between
the hard and soft momentum scales by design.  One would like to know
what happens when the back-reaction on the hard modes is not
completely negligible, which is probably the situation faced in real
experiments at finite energies. In this case one is naturally led to
consider instead the Wong-Yang-Mills (WYM) equations.  These equations
can be shown to reproduce the HL effective action in the weak-field
approximation~\cite{Kelly:1994ig,Kelly:1994dh,Blaizot:1999xk};
however, when solved fully they go beyond the HL approximation.

In this paper we present first real-time three-dimensional lattice
results obtained by solution of the WYM equations for coupled U(1) and
SU(2) particle-field systems.  For SU(2) gauge group, the WYM
equations describe the propagation of classical colored particles in a
colored background field in a framework that self-consistently
includes the deflection of the hard particles by the soft fields. In
practice, the equations are solved using the test-particle method and
we present an improved numerical algorithm which uses current smearing
to enable us to obtain convergent results with much fewer test
particles than the straightforward ``nearest grid point'' (point
particle) method. We focus here on simulations with large anisotropies
of the particles and, for SU(2), on initial field strengths beyond the
non-Abelian scale.

Our numerical results agree in some qualitative aspects with the HL
calculations in that we see a transfer of energy from particles to the
soft fields which saturates at late times. However, for the
simulations with large anisotropy presented here, the saturation seems
to stem from an ``avalanche'' of energy which has been dumped in soft
modes to higher momentum modes. Due to finite lattice spacing this
avalanche stops at the highest momentum modes of our lattice at which
time the rapid growth of the field energy density ends. As the lattice
spacing in our simulations is decreased we find that the amplitude at
which the field energy density saturates increases, indicating that
the mechanism for saturation is the population of hard lattice
modes. This is the main difference to the cascade which occurs in
hard-loop simulations with moderate anisotropies~\cite{Arnold:2005qs},
where field modes near the hard particle scale do not get populated
until parametrically large times. The behavior seen here might be
related to the chaoticity of three-dimensional classical Yang-Mills
fields~\cite{Matinyan:1986hc,Kawabe:1988jc,Biro:1993qc}. Additional
studies are necessary, however, before firm conclusions in this regard
can be drawn.

In Sec.~\ref{sec_WYMeqs} we present the Wong-Yang-Mills equations and 
the basics of the test particle method.  In Sec.~\ref{sec_NumMethods} 
we discuss the numerical methods employed. In Sec.~\ref{sec_results} 
we present results for U(1) and SU(2) gauge theories. In 
Sec.~\ref{sec_Discussion} we discuss the implications of our results 
and list open questions which remain.

\section{Wong-Yang-Mills equations}
\label{sec_WYMeqs}

In this paper we present solutions of the classical Vlasov transport
equation for hard gluons with non-Abelian color charge $q^a$ in the
collisionless approximation~\cite{Wong:1970fu,Heinz:1983nx},
\begin{equation}
 p^{\mu}[\partial_\mu - gq^aF^a_{\mu\nu}\partial^\nu_p
    - gf_{abc}A^b_\mu q^c\partial_{q^a}]f(x,p,q)=0~.   \label{Vlasov}
\end{equation}
Here, $f(t,\bm{x},\bm{p},q^a)$ denotes the single-particle phase space
distribution function and $g$ is the gauge coupling, which can take
arbitrary values at the classical level.

The Vlasov equation is coupled self-consistently to the Yang-Mills
equation for the soft gluon fields,
\begin{equation}
 D_\mu F^{\mu\nu} = J^\nu = g \int \frac{d^3p}{(2\pi)^3} dq \,q\,
 v^\nu f(t,\bm{x},\bm{p},q)~, \label{YM}
\end{equation}
with $v^\mu\equiv(1,\bm{p}/p)$. These equations reproduce the
``hard thermal loop'' effective action near
equilibrium~\cite{Kelly:1994ig,Kelly:1994dh,Blaizot:1999xk}. However, 
the full classical transport
theory~(\ref{Vlasov},\ref{YM}) also reproduces some higher $n$-point
vertices of the dimensionally reduced effective action for
static gluons~\cite{Laine:2001my} beyond the hard-loop approximation. The
back-reaction of the long-wavelength fields on the hard particles
(``bending'' of their trajectories) is, of course, taken into account.
This is important for understanding particle dynamics in strong
fields or for extremely anisotropic particle momentum distributions.

Eq.~(\ref{Vlasov}) can be solved numerically by replacing
the continuous single-particle distribution $f(\bm{x},\bm{p},q)$
by a large number of test particles:
\begin{equation}
 f(\bm{x},\bm{p},q) = \frac{1}{N_{\rm test}}\sum_i 
  \delta(\bm{x}-\bm{x}_i(t)) \,(2\pi)^3 \delta(\bm{p}-\bm{p}_i(t)) \,
   \delta(q^a-q_i^a(t))~,  \label{TestPartAnsatz}
\end{equation}
where $\bm{x}_i(t)$, $\bm{p}_i(t)$ and ${q}_i^a(t)$ are the
coordinates of an individual test particle and $N_{\rm test}$ denotes
the number of test-particles per physical particle. The {\sl
Ansatz}~(\ref{TestPartAnsatz}) leads to Wong's
equations~\cite{Wong:1970fu,Heinz:1983nx}
\begin{eqnarray}
\frac{d\bm{x}_i}{dt} &=& \bm{v}_i,\\ \frac{d\bm{p}_i}{dt} &=& g\,
q_i^a \,
\left( \bm{E}^a + \bm{v}_i \times \bm{B}^a \right),\label{pdot}\\
\frac{d\bm{q}_i}{dt} &=& ig\, v^{\mu}_i \, [ A_\mu, \bm{q}_i],\\ 
J^{a\,\nu} &=& \frac{g}{N_{\rm test}} \sum_i q_i^a \,v^\nu
\,\delta(\bm{x}-\bm{x}_i(t)).
\end{eqnarray}
for the $i$-th test particle.  The time evolution of
the Yang-Mills field can be followed by the standard Hamiltonian
method~\cite{Ambjorn:1990pu} in $A^0=0$ gauge.  Specific numerical
algorithms to solve the classical field equations coupled to particles
will be presented in the next section.

We use the following dimensionless lattice variables:
\begin{equation} \label{LatVars}
  \bm{E}_L^a = \frac{ga^2}{2}\bm{E}^a, \quad
  \bm{p}_L = \frac{a}{4}\bm{p},\quad
  Q_L^a = \frac{1}{2}q^a, \quad
  N_{\mathrm{test},L}  = \frac{N_\mathrm{test}}{g^2}~.
\end{equation}
Our lattice Hamiltonian\footnote{We employ the compact lattice
  formulation even for $U(1)$ gauge group.} is then
\begin{equation}
 H_L = \frac{1}{2}\sum_{i}\bm{E}_{i,L}^{a\,2}
       + \frac{1}{2}\sum_{\Box} \left( N_c - {\rm Re}\;\Tr \,U_{\Box}\right)
       + \frac{1}{N_{\mathrm{test},L}} \sum_{j}|\bm{p}_{j,L}|,
\end{equation}
which is related to the physical Hamiltonian by $H = (4/g^2a)\, H_L$.
To convert lattice variables to physical units we fix the length of
the lattice to $L=5$~fm which then determines the physical scale for
the lattice spacing $a$. All other dimensionful quantities can then be
converted to physical units using eqs.~(\ref{LatVars}). In other
words, simulations on lattices with increasing number of sites
correspond to a fixed infrared cutoff, while the ultraviolet cutoff
increases in proportion to the number of sites. In lattice units,
$p_{h,L}$ should not be much smaller than 1; otherwise, the particle
and field modes would overlap significantly.

For most of the results presented here the initial phase-space
distribution of hard gluons is taken to be a (parity-invariant) planar
momentum distribution:
\begin{equation}
   f(\bm{p}) = n_g \left(\frac{2\pi}{p_h}\right)^2\delta(p_z)
              \exp(-p_T/p_h)~, \label{AnisoDistrib}
\end{equation}
with $p_T=\sqrt{p_x^2+p_y^2}$ and $n_g$ the number density of hard
gluons (summed over two helicities and, for $SU(N_c)$ gauge group,
also over $N_c^2-1$ color states). This represents a quasi-thermal
distribution in two dimensions, with average momentum equal to
$p_h$. At the initial time $t=0$ we randomly sample $N_p = N_{\rm
test}\, n_g\, a^3$ particles from the
distribution~(\ref{AnisoDistrib}) at each cell of the
lattice.\footnote{When $N_p$ is not very large it is useful to ensure
explicitly that the sum of particle momenta in each cell vanishes, for
example by adjusting the momentum of the last particle accordingly.}

We shall also present data for distributions with more moderate or
vanishing anisotropy for numerical checks, where indicated. Instead of the
number density $n_g$ of hard gluons it is also customary to quote the
assumed value for the mass scale
\begin{equation}
  m^2_{\infty} = g^2N_c \int\frac{d^3p}{(2\pi)^3} \frac{f(\bm{p})}{|\bm{p}|}
             \sim g^2N_c \, \frac{n_g}{p_h}~.
\end{equation}
(For the particular distribution~(\ref{AnisoDistrib}) the latter is
actually an equality.) This quantity sets the scale for the growth
rate of unstable field modes in the linear approximation.

The initial field amplitudes are sampled from a Gaussian distribution
with a width tuned to a given initial energy density:
\begin{equation}
 \langle A^a_i(\bm{x})A^b_j(\bm{y}) \rangle
    = \frac{4\mu^2}{g^2} \delta_{ij}\delta^{ab}\delta(\bm{x}-\bm{y})~,
\end{equation}
and $\bm{E}=-\dot{\bm{A}}=0$. Gauss's law then implies that the local
charge density at time $t=0$ vanishes. We ensure that any particular
initial condition satisfies {\em exact} local charge neutrality. In
our $U(1)$ simulations we monitor the time evolution of Gauss's law as
a check of the numerical accuracy. Our charge smearing algorithm for
$SU(2)$, on the other hand, explicitly exploits (covariant) current
conservation and hence Gauss's law is satisfied exactly by
construction.  For $U(1)$ simulations the results of single simulation
runs are shown but for our $SU(2)$ simulations observables are averaged 
over several initial field and particle configurations.  Where shown
error bars indicate the root-mean-square variation over this ensemble.

\section{Numerical methods}
\label{sec_NumMethods}

\subsection{NGP method for non-Abelian gauge theories}
We first summarize the method developed in Ref.~\cite{Hu:1996sf,Moore:1997sn}
where zero-order weighting is used (i.e., point particles). In that
approach, one simply counts the number of particles within a cell to
determine the charge density, which is why it is called the
nearest-grid-point (NGP) method.  A current is generated only when a
particle crosses a cell boundary from $i$ to $i+1$. This induces a
current on that link which is $J(i)=Q\delta(t -
t_\mathrm{cross})/N_\mathrm{test}$.  In order to satisfy the lattice
covariant continuity equation,
\begin{equation}
 \dot{\rho_i} = \sum_x U^\dagger_x(i-x)J_x(i-x)U_x(i-x) - J_x(i)~,
\end{equation}
the charge must be parallel transported:
\begin{equation}
  Q(i+1) = U^\dagger_x(i)Q(i)U_x(i)~.
\end{equation}
The rotation of the particle's momentum by the magnetic field can be
updated continuously but its magnitude changes only when it crosses a link.
The momentum ``kick'' can be determined by energy conservation:
\begin{equation}
  |\bm{p}_\mathrm{ini}| + \frac{N_\mathrm{test}}{2}\bm{E}_\mathrm{ini}^2
  =
  |\bm{p}_\mathrm{fin}| + \frac{N_\mathrm{test}}{2}
       (\bm{E}_\mathrm{ini} - J )^2~.
\end{equation}
Substituting $J={Q}/{N_\mathrm{test}}$ and assuming that the
particle crosses the cell boundary in $x$-direction, one obtains
\begin{equation}
  |\bm{p}_\mathrm{fin}| = |\bm{p}_\mathrm{ini}|
	- E_{x,\mathrm{ini}}Q + Q^2/(2N_\mathrm{test})~.
\end{equation}
This way, the total energy
\begin{equation}
 E_\mathrm{tot} = \frac{1}{2}\sum_{\mathrm{lattice}}( \bm{E}^2 + \bm{B}^2 )
                + \frac{1}{N_\mathrm{test}}\sum_i |\bm{p}_i|
\end{equation}
is conserved.

In principle, the NGP method also requires time ordering of the link
crossings by particles~\cite{Hu:1996sf,Moore:1997sn} in every time
step, which becomes quite time consuming when the number of
test-particles is large. However, we have found that at least for our
present purposes time-ordering does not affect the results
significantly and so will be ignored in what follows. 

This method was applied 
in~\cite{Dumitru:2005gp,Dumitru:2005hj,Nara:2005fr} to study a 
simplified situation with fields and particles living on a one-dimensional 
spatial lattice. Physically, this corresponds to a 
transversally homogeneous system (translational invariance in $x$ and 
$y$ directions). For such 1d-3v simulations$\,$\footnote{Fields and charge 
densities fluctuate only in $z$-direction but particle velocities can 
point in any direction.} the current can be made sufficiently smooth 
by employing enough test-particles. 
Ref.~\cite{Dumitru:2005gp,Dumitru:2005hj,Nara:2005fr} found that the 
currents were sufficiently smooth when the number of particles per 
lattice site, $N_p = N_{\rm test}\, n_g\, a^3$, was on the order of a 
few hundred to a thousand.

\begin{figure}[ht]
\includegraphics[width=3.0in]{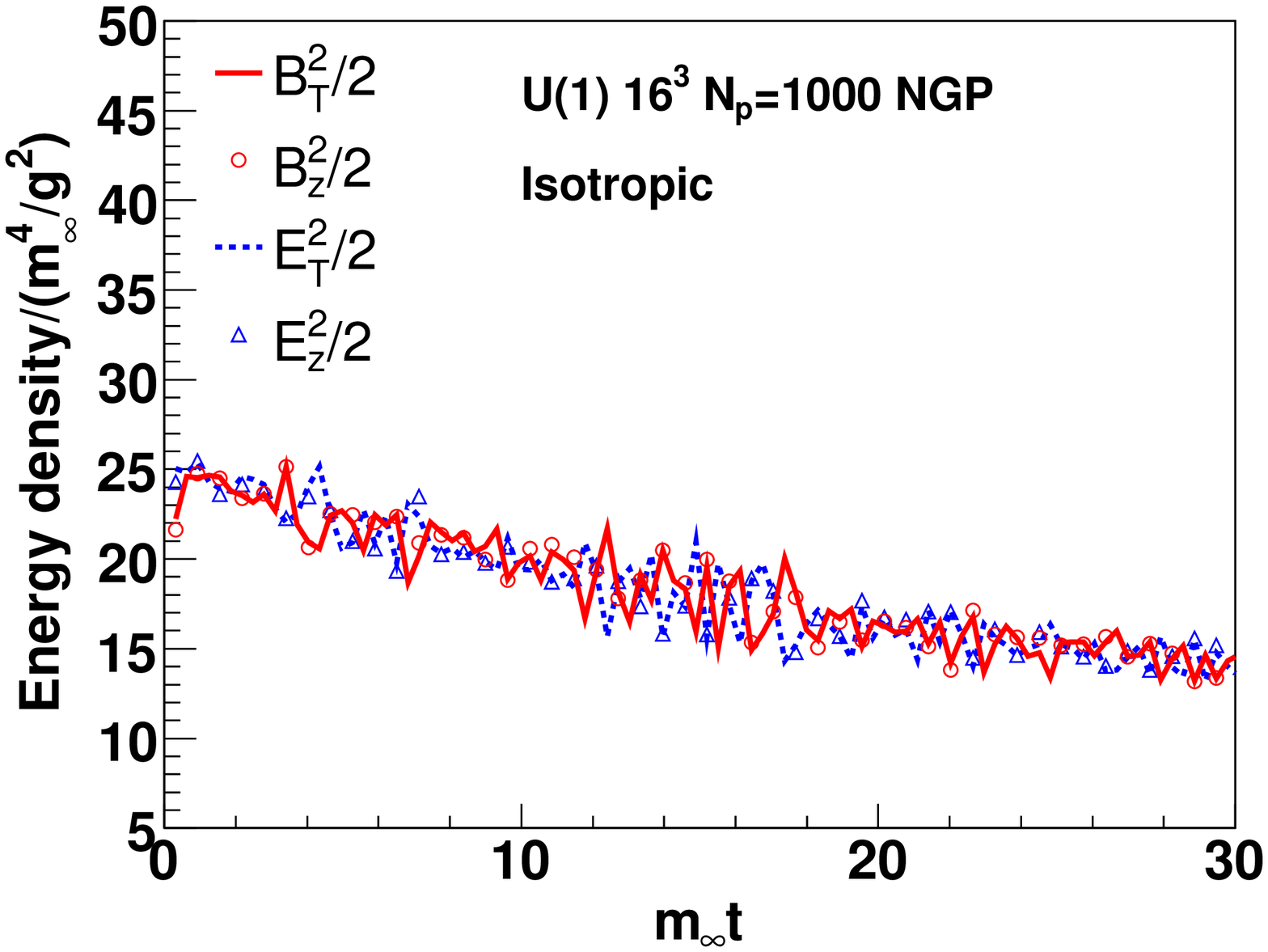}
\includegraphics[width=3.0in]{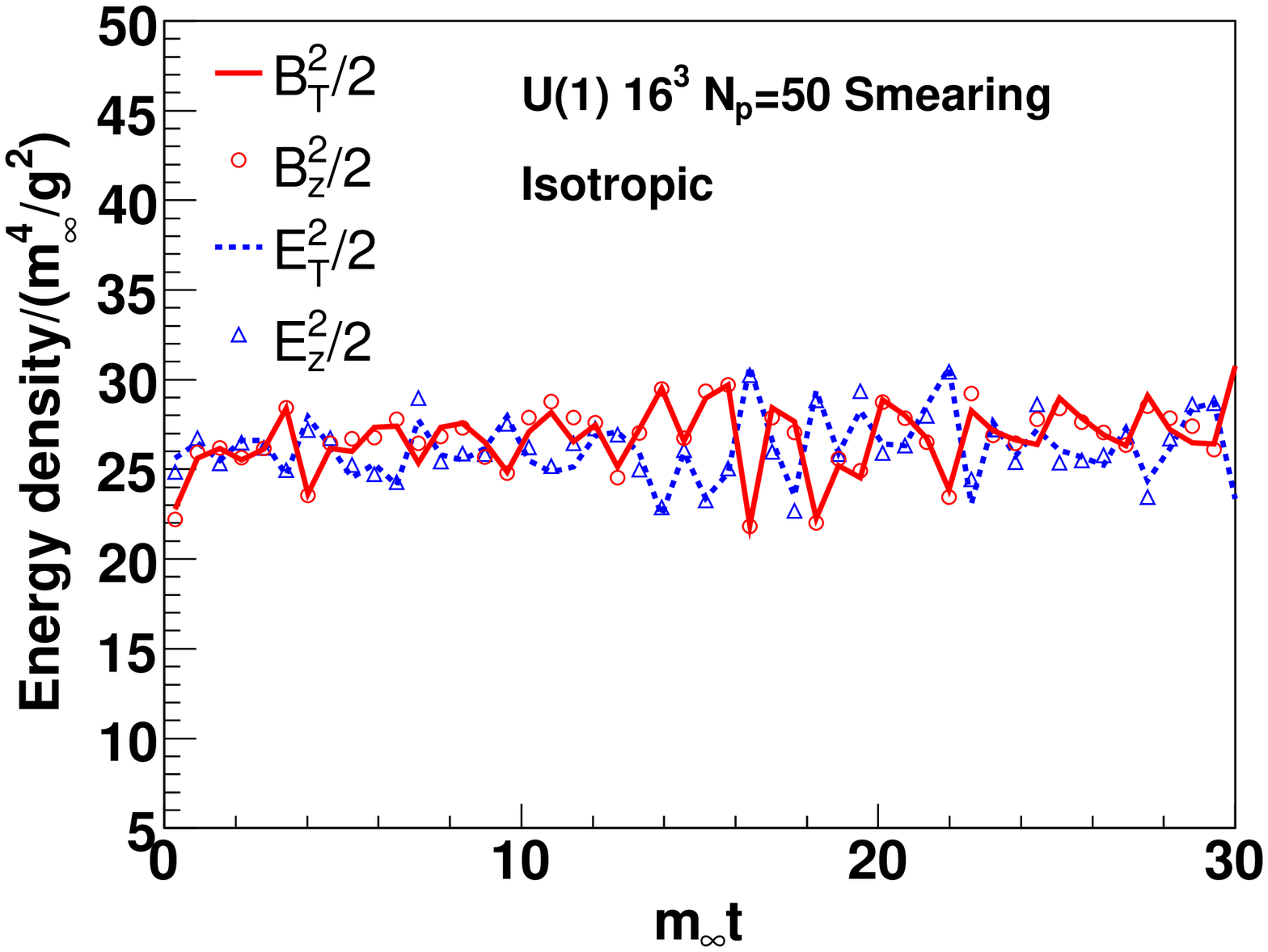}
\caption{ Time evolution of the magnetic and electric energy densities
for $U(1)$ gauge group and isotropic particle momentum distribution using
the nearest-grid-point method (left) and smeared-particles (right). 
}
\label{fig:isoNGP}
\end{figure}

Full 3d-3v simulations, however, require improved algorithms since 3d 
lattices have many more sites and the NGP method becomes 
computationally too expensive. In Fig.~\ref{fig:isoNGP} (left), the 
time evolution of the field energy densities on a $16^3$ lattice with 
$N_p=1000$ test-particles per site are shown. Despite the large number 
of particles, there seems to be an anomalous damping of the fields
with the NGP method. We naively expect that to achieve the same 
accuracy as for a one-dimensional lattice, $N_p$ would have to 
increase exponentially with the number of spatial dimensions: 
$N_p^{3d}\sim (N_p^{1d})^3$. This makes multi-dimensional NGP 
simulations practically impossible and forces one to consider methods 
for smoothing particle densities and currents.  

\subsection{Current smearing in electromagnetic PIC simulations}

In Abelian plasmas it is common to employ smoothed currents for
particle-in-cell (PIC) simulations~\cite{Hockney:1985,Birdsall:1985} in order to avoid
numerical noise.  The charge density at site $(i,j,k)$ is obtained by
smearing$\,$\footnote{We assume $\Delta x=\Delta y=\Delta z=1$.}
\begin{equation}
 \rho(i,j,k) = e \sum^{N_p}_{n=1}  S_i(x_n) S_j(y_n) S_k(z_n)~.
\end{equation}
Here, $S$ is a form factor. For example, the first-order form factor
(shape-factor) is
\begin{equation}
  S^1_i(\xi) = \left\{
     \begin{array}{ll}
      1 - |\xi - i| & \mathrm{for}\, |\xi - i| \leq 1,\\
       0            & \mathrm{for}\, |\xi - i| > 1
       \end{array}
       \right.
\end{equation}
However, it is well known that if we naively define the current
density at each grid point as
\begin{equation}
  J_\alpha(i,j,k) = e \sum^{N_p}_{n=1} v_{\alpha,n}
                    S_i(x_n) S_j(y_n) S_k(z_n), \quad \alpha=x,y,z,
		   \label{eq:current}
\end{equation}
then the continuity equation is not satisfied exactly~\cite{Hockney:1985,Birdsall:1985},
requiring us to correct the electric field by solving the Poisson
equation. 

To avoid solving Poisson's equation in each time step,
several numerical techniques for solving the continuity equation
have been developed~\cite{Eastwood:1991,Eastwood:1995,Buneman:1992,Esirkepov:2001,Umeda:2003}.
For simplicity, we consider the 2d problem to explain
how to satisfy local charge conservation in electromagnetism.
We consider a particle moving from $(x_1, y_1)$ to $(x_2, y_2)$
during a time step $\Delta t$, i.e.
$x_2= x_1 + v_x\Delta t, \quad y_2= y_1 + v_y\Delta t$.
We also restrict ourselves to the case in which
$(x_2,y_2)$  belongs to the same lattice site $(i,j)$.
Charge conservation of the assigned current densities
is realized with the procedure described by Eastwood~\cite{Eastwood:1991},
\begin{eqnarray}
 J_x(i,j) &=& F_x(1-W_y), \qquad J_x(i,j+1) = F_xW_y~, \nonumber\\
 J_y(i,j) &=& F_y(1-W_x), \qquad J_y(i+1,j) = F_yW_x~, \label{eq:smCurrent}
\end{eqnarray}
where $(F_x,F_y) \equiv \bm{F}$ represents the charge flux
\begin{equation}
 F_x = e\frac{x_2-x_1}{\Delta t}~, \qquad
 F_y = e\frac{y_2-y_1}{\Delta t}~.
\end{equation}
$W$ represents the first-order shape-factor corresponding to the
linear weighting function defined at the midpoint between
the starting point $(x_1,y_1)$ and the end point $(x_2,y_2)$
\begin{equation}
  W_x = \frac{x_1+x_2}{2}-i~,\qquad
  W_y = \frac{y_1+y_2}{2}-j~.
\end{equation}
One can easily show that the lattice continuity equation
\begin{equation}
  J_x(i,j)-J_x(i-1,j) + J_y(i,j)-J_y(i,j-1)
   = \frac{\rho^t(i,j)-\rho^{t+\Delta t}(i,j)}{\Delta t} \, ,
\end{equation}
is satisfied with the definition~(\ref{eq:smCurrent}) of the current.
When $(x_2,y_2)$ is a different lattice site than the original $(i,j)$
one, we use the `Zigzag scheme' developed in Ref.~\cite{Umeda:2003}.
Much of the numerical noise is eliminated by such linear interpolation. 
Note that our current may in general be expressed as
\begin{equation}
  J_\alpha(i,j,k) = e \sum^{N_p}_{n=1} v_{\alpha,n}
                    S^0_\alpha(x_n) S^1_j(y_n) S^1_k(z_n), 
		    \quad \alpha=x,y,z, \quad \alpha \neq j,k,
		   \label{eq:current2}
\end{equation}
which differs from~(\ref{eq:current}) since $S^0_\alpha =
dS^1_\alpha/dx_\alpha$; $S^0_\alpha$ is the ``form factor'' of the NGP
method, i.e.\ $S^0_\alpha= 1$ if the particle is in the corresponding
lattice site, otherwise it is zero.

Finally, we note that the electromagnetic forces should be smeared in
a similar way when a particle momentum is
updated~\cite{Eastwood:1991,Eastwood:1995}. Consider the time
derivative of the total energy:
\begin{equation}
 \frac{dE}{dt} = -\sum_\text{lattice} \bm{E}\cdot \bm{J}
             + \sum_\text{particles} q_i \, \bm{E}(\bm{x}_i(t))\cdot \bm{v}_i.
\end{equation}
It is clear that the interpolation function for $\bm{E}$ should be the
same as that for $\bm{J}$ in order to achieve good energy conservation
in the simulation.  The electric field $\bm{E}(\bm{x})$ at the particle
position $\bm{x}$ is then obtained from
\begin{equation} \label{Efieldx}
 E_\alpha(\bm{x}) = \sum_{\text{lattice}}
         S^0_\alpha S^1_\beta S^1_\gamma E_\alpha(i,j,k)~,
\end{equation}
while the magnetic field is given by
\begin{equation} \label{Bfieldx}
 B_\alpha(\bm{x}) = \sum_{\text{lattice}}
         S^1_\alpha S^0_\beta S^0_\gamma B_\alpha(i,j,k)~.
\end{equation}
This is motivated by the relations $\bm{E} =-\dot{\bm{A}}$ and
$ \bm{B} = \nabla \times \bm{A}$.
The momentum update itself is explained in appendix~\ref{appendixMomUpdate}.

In Fig.~\ref{fig:isoNGP} (right) we show the $U(1)$ energy densities
resulting from an isotropic run using $N_p=50$ smeared test-particles
per site. As can be seen from this figure the smeared-particle result
does not suffer from the anomalous damping seen with the NGP method.
More importantly, this result was obtained using many fewer particles
than would be required to obtain a stable result with the NGP method.
Therefore the smeared-particle method makes three-dimensional
particle-in-cell (PIC) simulations possible in practice.

\begin{figure}[ht]
\includegraphics[width=3.0in]{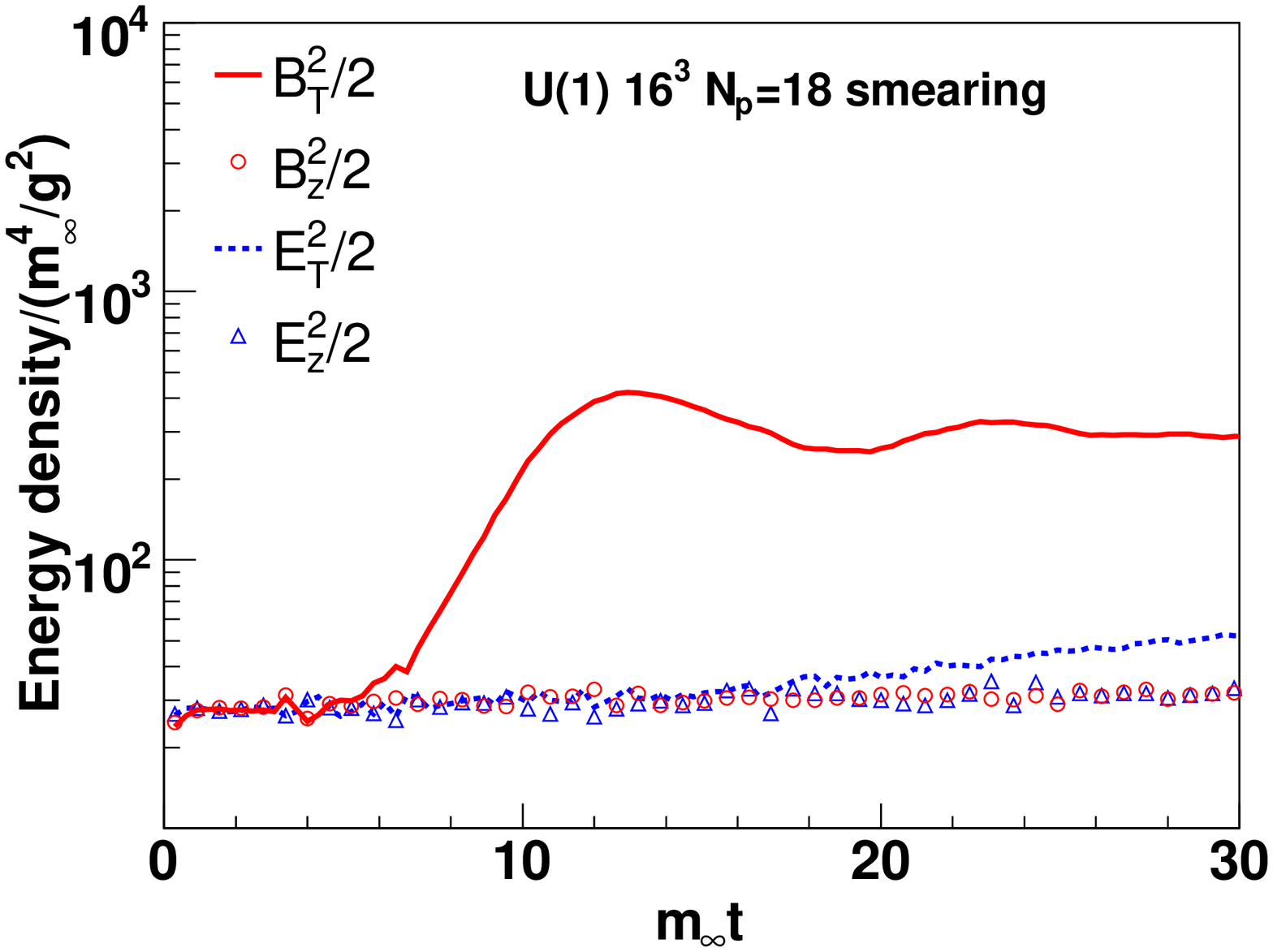}
\includegraphics[width=3.0in]{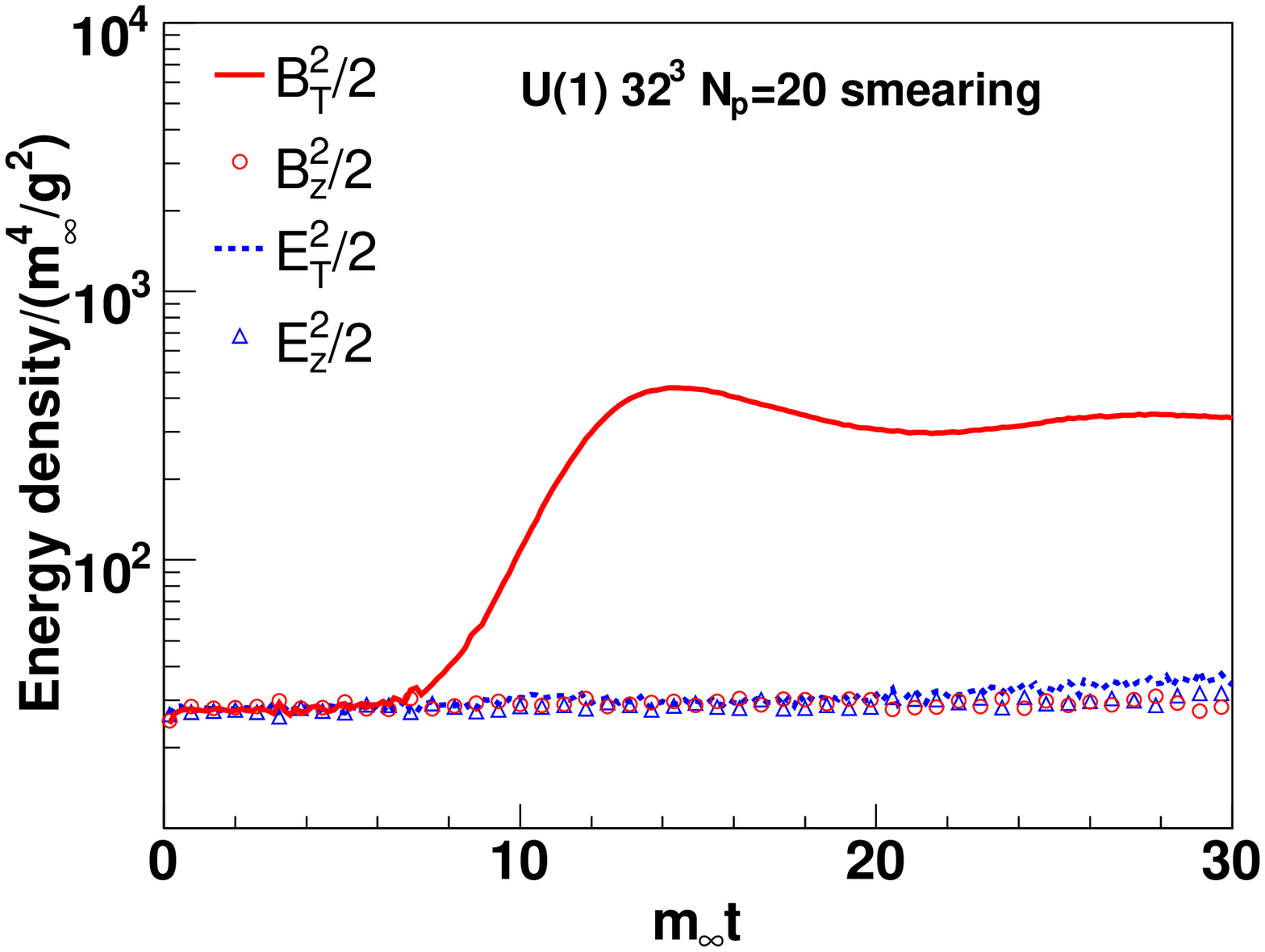}
\includegraphics[width=3.0in]{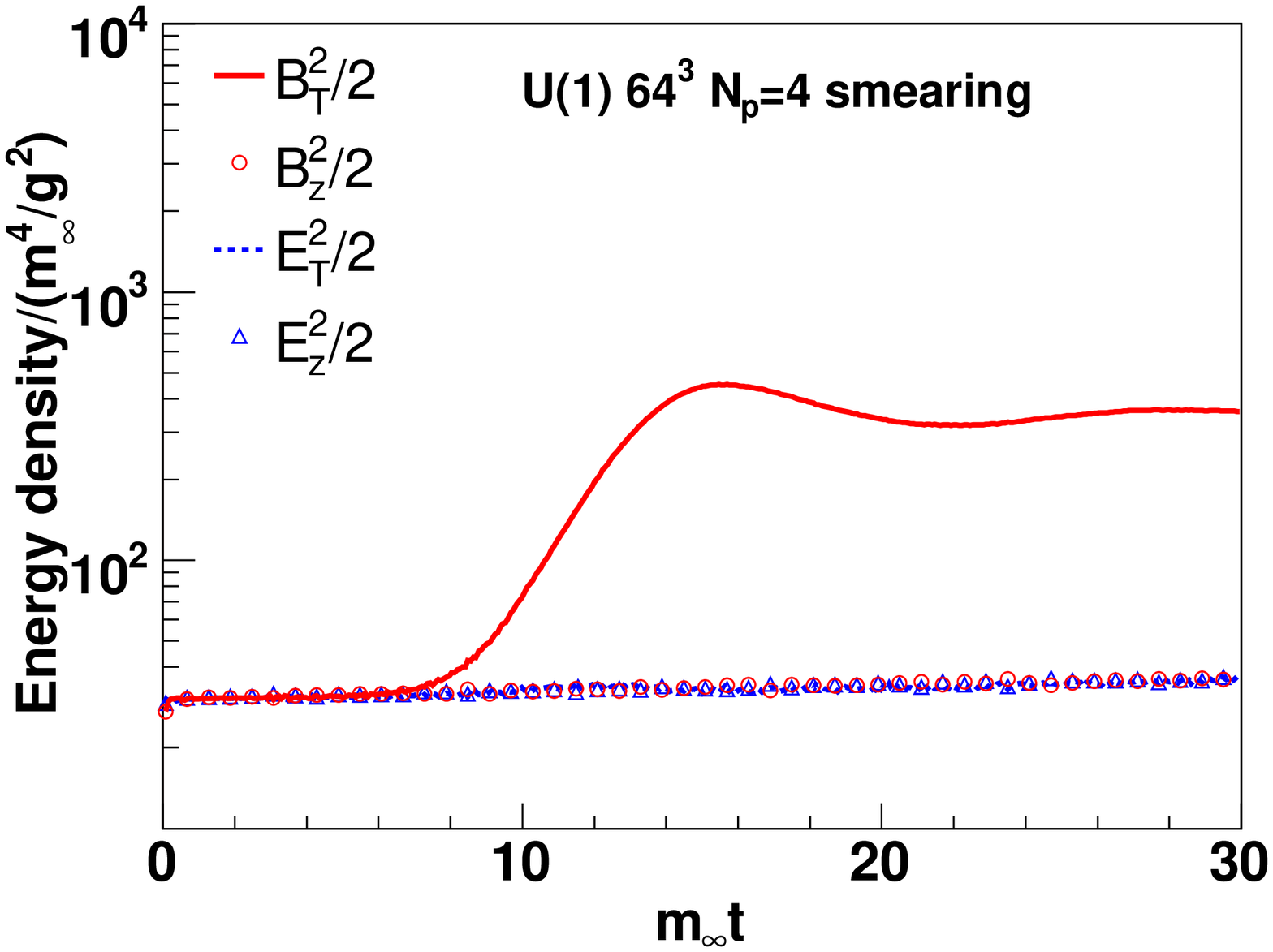}
\includegraphics[width=3.0in]{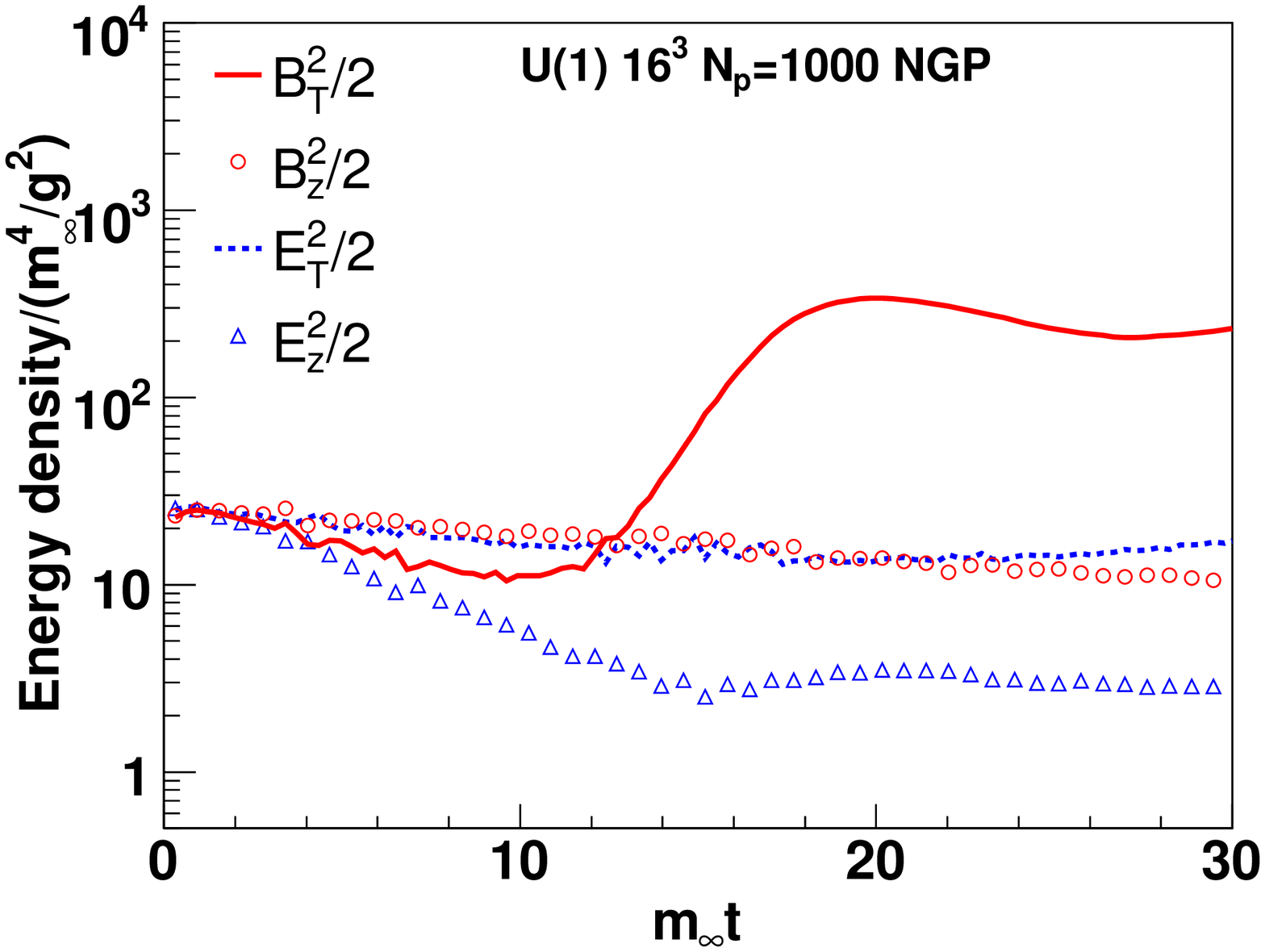}
\caption{ Time evolution of the field energy densities for $U(1)$
gauge group and anisotropic particle momentum distribution
(\ref{AnisoDistrib})
on $16^3$, $32^3$ and $64^3$ lattices.  Simulation parameters are $L=5$~fm,
$p_h=16$~GeV, $g^2\,n_g=20$~fm$^{-3}$, $m_\infty=0.1$~GeV.}
\label{fig:fieldWu1}
\end{figure}

Fig.~\ref{fig:fieldWu1} shows the $U(1)$  field energy densities as a function
of time for the anisotropic initial particle momentum distribution using
both smeared particles and the NGP method.  These simulations illustrate 
the improved accuracy of the smearing method as
compared to the NGP method. Moreover, thanks to the smaller number of
test-particles required by the smearing method we are now able to also
run simulations on larger lattices. For these initial conditions, we
observe an instability to transverse magnetic fields while all other
field amplitudes are essentially constant over the depicted time
interval.  We also note that the evolution of the fields is
essentially independent of the lattice spacing, indicating that only
soft field modes are important for the dynamics of the instability in
the Abelian theory. Below, we shall find that the situation is
different for non-Abelian fields.

The growing fields deflect the particles if their momenta are not
asymptotically large and as a consequence they aquire non-vanishing
longitudinal momenta $p_z$ (c.f.\ section~\ref{sec_results}).  We
remark that PIC methods are not well suited for simulations of the
extreme weak-field limit, where the particles propagate on nearly
straight-line trajectories for a very long time.\footnote{This is
better done within the hard-loop approximation.}  This is not due to
physical but technical limitations: the weaker the fields, the more
accurate the particle current $J^\nu$ needs to be (for example with
respect to cancellations of positively and negatively charged
particles etc.), as it represents a source-term in the field
equations. High-precision currents, in turn, can only be achieved by
employing a large number of test particles, driving up the
computational time. In practice, the weakest fields for which
simulations were feasible correspond to initial energy densities of
about $0.1\,m^4_\infty/g^2$ for $U(1)$ and $\sim10\,m^4_\infty/g^2$
for $SU(2)$ gauge group, respectively.  The natural regime of
applicability of PIC methods is when the separation between hard and
soft degrees of freedom is not asymptotically large.

\subsection{PIC simulations in non-Abelian gauge theories (CPIC)}
\label{nonAbelianPIC}
A straightforward extension of the above-mentioned smearing method to
the non-Abelian case would be to define the current as
\begin{eqnarray}
 J_x(i,j) &=& Q\frac{x_2-x_1}{\Delta t}(1-W_y),
         \qquad J_x(i,j+1) = Q_{y}\frac{x_2-x_1}{\Delta t}W_y~, \\
 J_y(i,j) &=& Q\frac{y_2-y_1}{\Delta t}(1-W_x), \qquad 
  J_y(i+1,j) = Q_{x}\frac{y_2-y_1}{\Delta t}W_x~, 
\end{eqnarray}
where we define the parallel transport of the charge in 2d as$\,$\footnote{We 
omit the term $[A_z,J_z]$ here, because this term does
not appear in 3d and because it does not matter for the present
discussion.}
\begin{equation}
 Q_x \equiv U_x^\dagger(i,j)QU_x(i,j),\qquad
 Q_y  \equiv  U_y^\dagger(i,j)QU_y(i,j)~.
\end{equation}
One can easily check that
this satifies the lattice covariant continuity equation,
\begin{equation}
 \dot{\rho}(i,j) = \sum_x U^\dagger_x(i-x)J_x(i-x)U_x(i-x) - J_x(i,j),
\end{equation}
for sites $(i,j), (i+1,j), (i,j+1)$:
\begin{eqnarray}
\dot{\rho}(i,j) &=& J_x(i,j) + J_y(i,j), \label{eq:ch1}\\
\dot{\rho}(i+1,j)&=&
     U^\dagger_x(i,j)J_x(i,j)U_x(i,j) - J_y(i+1,j),\label{eq:ch2}\\
\dot{\rho}(i,j+1) &=&
    U^\dagger_y(i,j)J_y(i,j)U_y(i,j) - J_x(i,j+1), \label{eq:ch3}
\end{eqnarray}
Eqs.~(\ref{eq:ch1}), (\ref{eq:ch2}), (\ref{eq:ch3}) are
consistent with the following definitions of the charge densities:
\begin{eqnarray}
 \rho(i,j) &=& Q(1-x)(1-y) ~,\\
 \rho(i,j+1) &=& Q_y(1-x)y ~,\\
 \rho(i+1,j) &=& Q_xx(1-y)~.
\end{eqnarray}
%
%
However, since a particle's color charge depends on its path, so does
$\rho(i+1,j+1)$ and we are not able to calculate it from the charge
distribution itself. Rather, we directly employ covariant current
conservation to determine the increment of charge at site $(i+1,j+1)$
within the time-step.  This way, we can satisfy Gauss's law in the
non-Abelian case. The forces acting on a particle are determined by
analogy to~(\ref{Efieldx},\ref{Bfieldx}), except that the charges
at neighboring sites is given by the expressions above.

Finally, we have to check that $\Tr(Q^2)$ is conserved by this smearing
method. This is true when the lattice spacing $a$ is small, as the
total charge of a particle is given by
\begin{equation}
 Q_0 = Q(1-x)(1-y) + Q_xx(1-y) + Q_y(1-x)y + [a_pQ_{xy} + (1-a_p)Q_{yx}]xy~,
\end{equation}
where the $a_p$ depend on the path of a particle and
$Q_{xy}=U^\dagger_x(i,j+1)Q_yU_x(i,j+1)$,
$Q_{yx}=U^\dagger_y(i+1,j)Q_xU_y(i+1,j)$.
If we require that $\Tr(Q_0^2)$ be constant, then the
cross terms, for example $\Tr(QQ_x)$, have to vanish.
This is true when $a$ is small, because $\Tr(Q[A,Q])=0$:
\begin{equation}
 \Tr(QQ_x) = \Tr(Q(Q + iga[A_x,Q] + {\cal O}(a^2))) = \Tr(Q^2) +{\cal O}(a^2).
\end{equation}


\section{Results}  
\label{sec_results}

\subsection{Electrodynamics}

\begin{figure}[htb]
\includegraphics[width=3.0in]{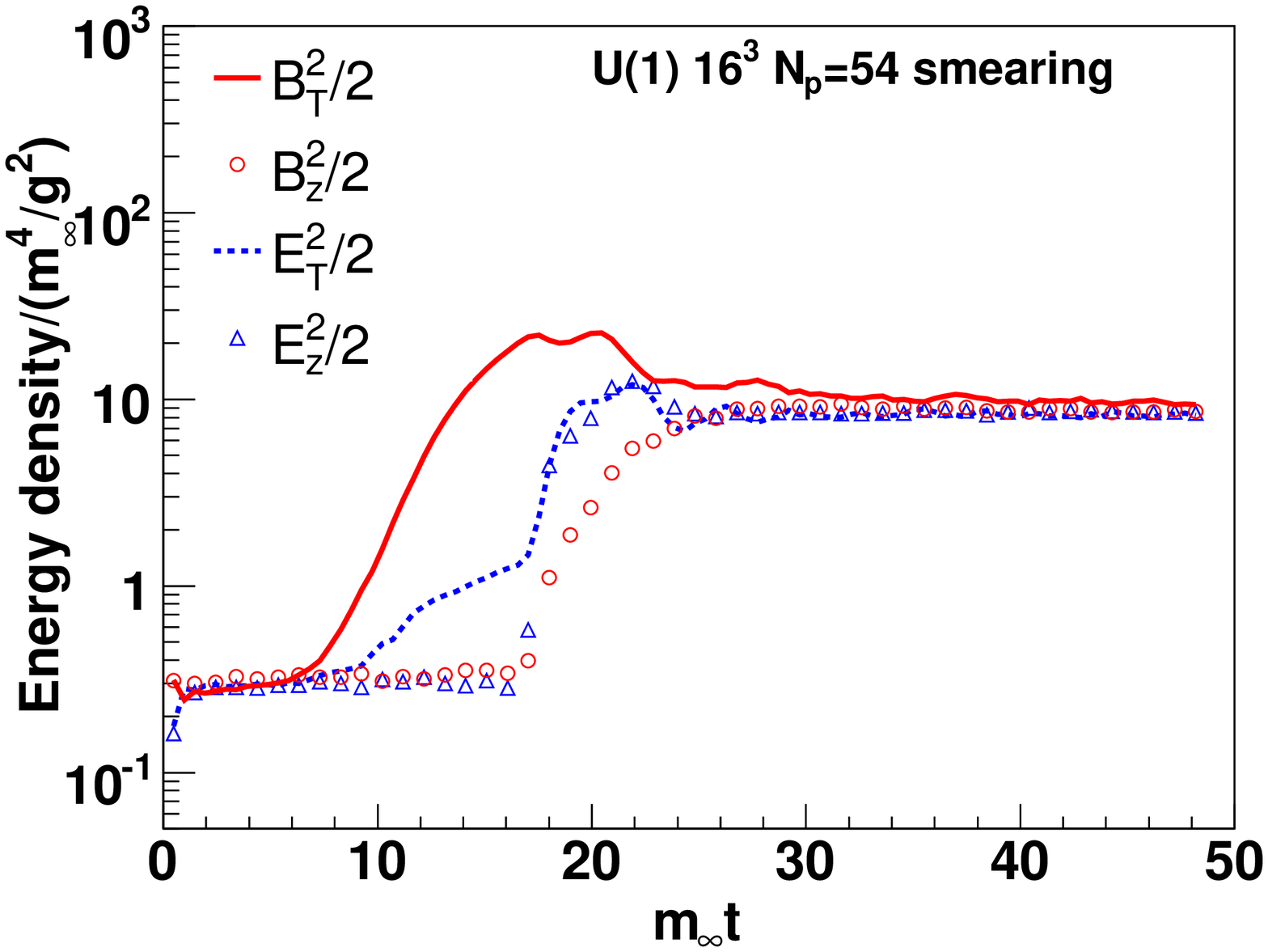}
\includegraphics[width=3.0in]{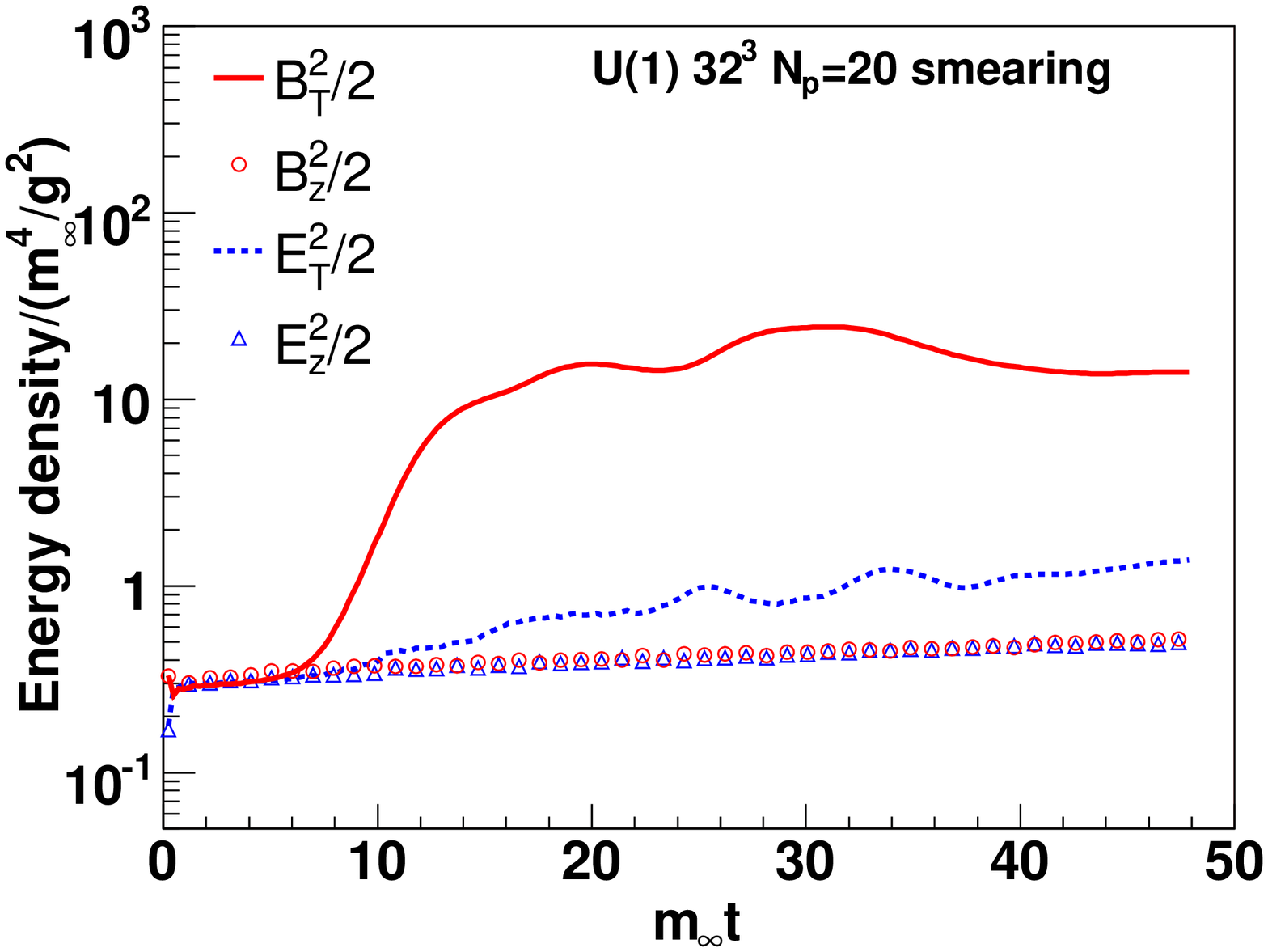}
\includegraphics[width=3.0in]{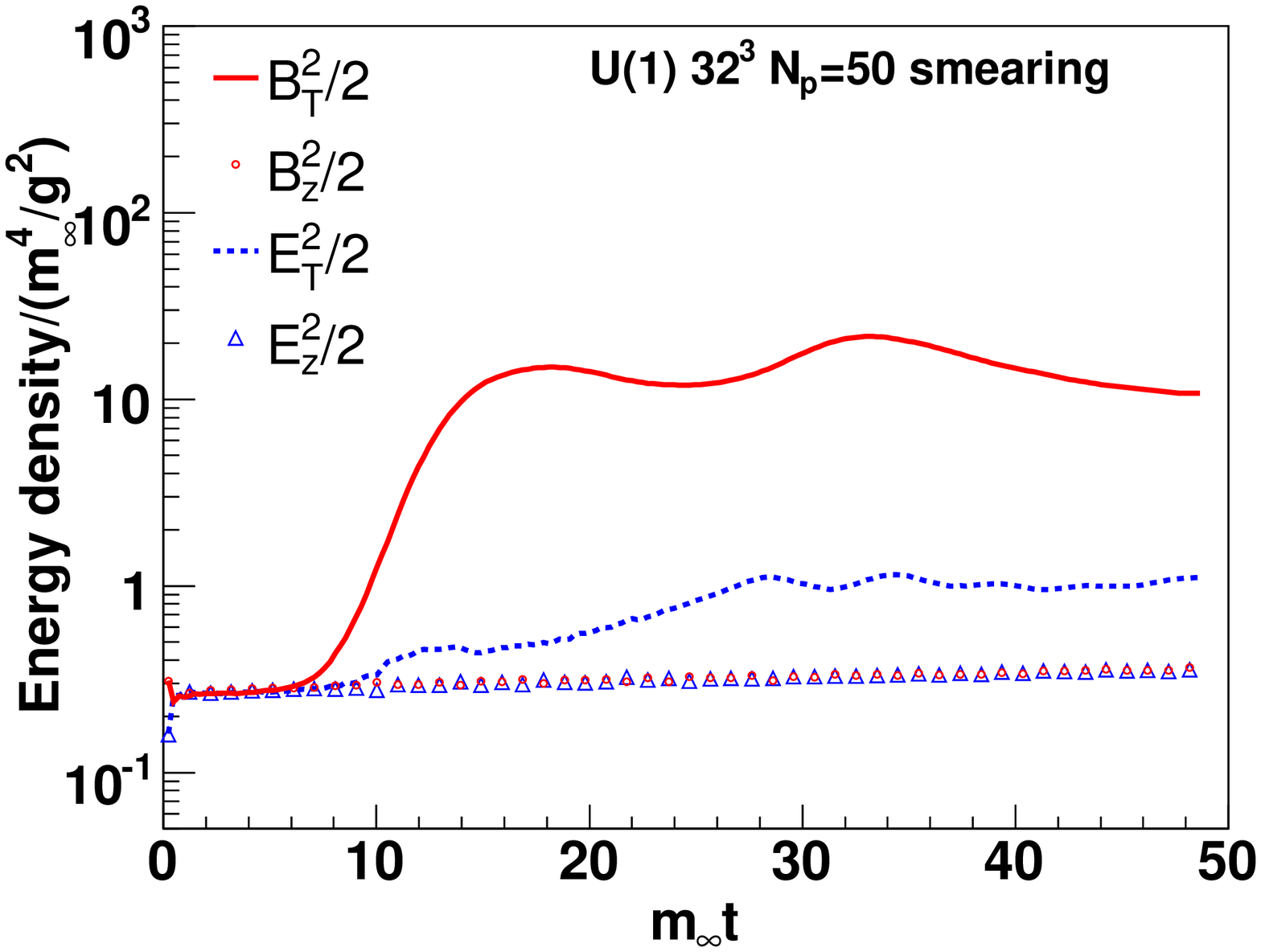}
\includegraphics[width=3.0in]{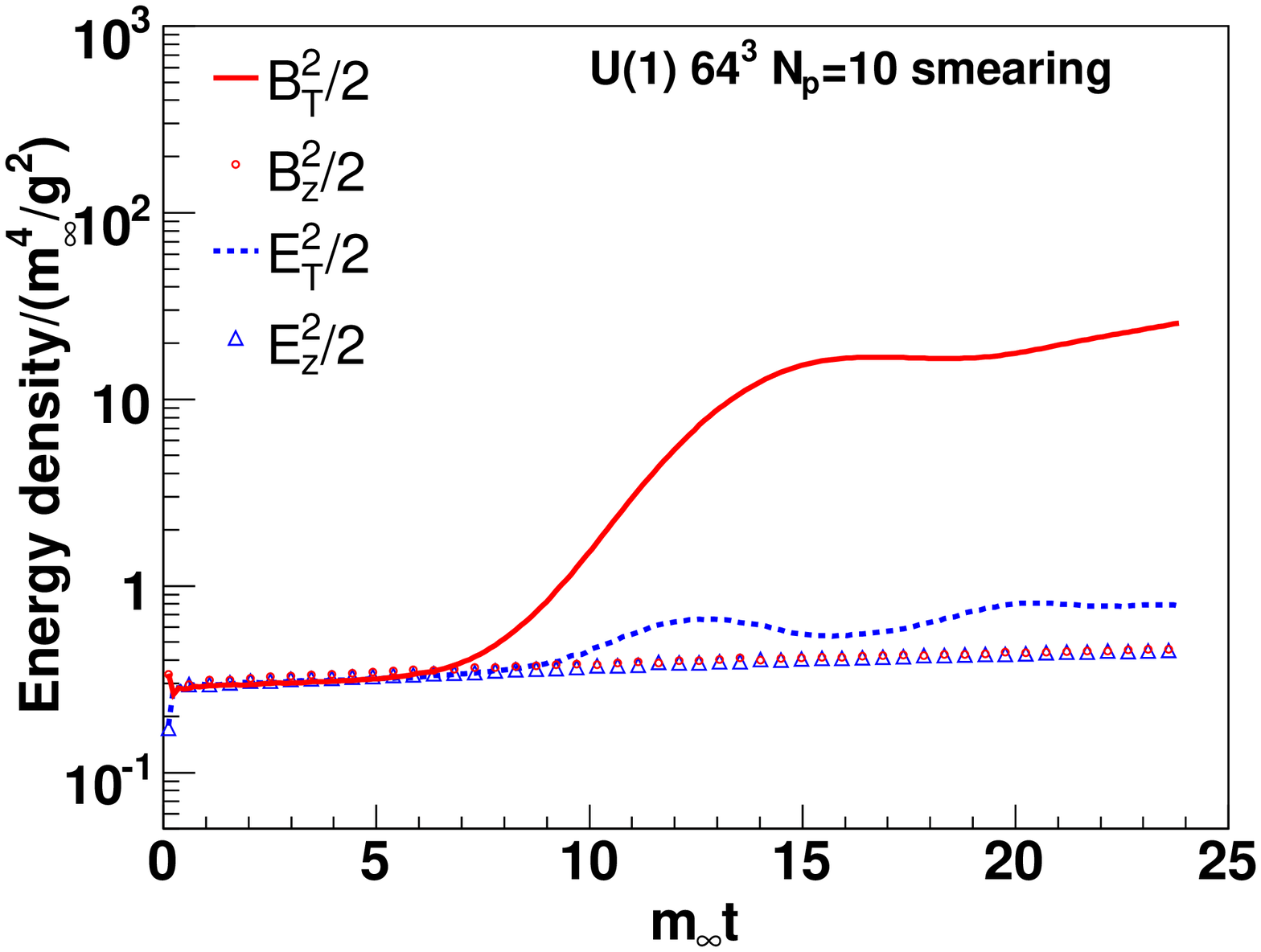}
\caption{Time evolution of the field energy densities for $U(1)$
gauge group and anisotropic particle momentum distribution on various
lattices.  Simulation parameters in physical units are $L=5$~fm,
$p_h=8$~GeV, $g^2 \,n_g=100$~fm$^{-3}$, $m_\infty=0.3$~GeV.}
\label{fig:fieldWu1_weak}
\end{figure}
In Fig.~\ref{fig:fieldWu1_weak} we show simulation results for
``weak'' fields with initial energy density on the order of
$0.1\,m_\infty^4/g^2$.  For this case, we find that not only the
transverse magnetic but also the transverse electric fields grow,
albeit at a slower rate.\footnote{There is little difference to 1d-3v
simulations, compare for example to Fig.~1 of 
ref.~\cite{Dumitru:2005gp,Dumitru:2005hj,Nara:2005fr}.} Note
that here, and throughout the manuscript, squared transverse field
components denote {\em averages} of squares of $x$- and
$y$-components: $B_T^2=(B_x^2+B_y^2)/2$, $E_T^2 = (E_x^2 +
E_y^2)/2$. For the $16^3$ lattice there is a very sudden growth of all
field components at $t\simeq16\,m^{-1}_\infty$ which we deem
unphysical since we do not expect equipartitioning of transverse and
longitudinal fields; this is an example for the problems which arise
in PIC simulations when the fields are very weak and the currents are
not sufficiently stable\footnote{Also, on coarse lattices non-linear
  terms in the Maxwell equations, which arise in the compact lattice
  formulation of $U(1)$, may play a role. This is probably not the
  case here, however, since qualitatively correct results were
  obtained on the same $16^3$ lattice for {\em higher} field strength,
Fig.~\ref{fig:fieldWu1}.}. Indeed, this numerical instability disappears
on larger lattices with more test-particles. The longitudinal field
components are now essentially constant and only the transverse
components grow.  The time where the instability for the transverse
magnetic fields sets in, as well as their growth rate and saturation
point is quite similar for the $32^3$ and $64^3$ lattices.

\begin{figure}[htb]
\includegraphics[width=4.0in]{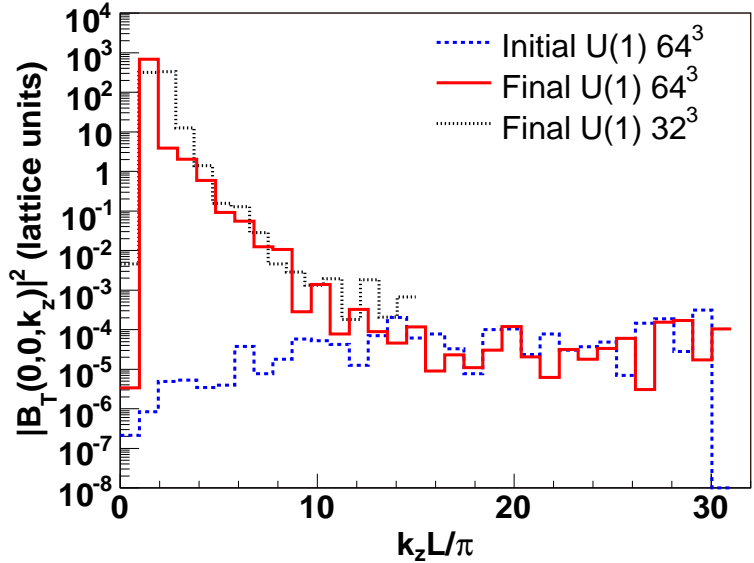}
\caption{Fourier transform of the tranverse magnetic field from
  Fig.~\ref{fig:fieldWu1_weak}.}
\label{fig:FT_fieldWu1_weak}
\end{figure}
The slope of the exponential growth of $B_T^2$ is about half that
predicted by the hard-loop approximation ($=\sqrt{2}\,m_\infty$ for
the field energy density) which indicates that non-linear effects are
not completely negligible. This is also apparent from the much slower
growth of $E_T$ as compared to $B_T$ and in the Fourier transform of
the transverse magnetic field shown in Fig.~\ref{fig:FT_fieldWu1_weak}
which illustrates the spectrum of unstable modes. In the linear
approximation, the spectrum of unstable modes for an extreme particle
anisotropy extends to $k_{\rm max}\sim m_\infty/\theta$, where
$\theta$ denotes the typical angular width of the particle momentum
distribution.  For our initial condition~(\ref{AnisoDistrib})
$\theta=0$ at $t=0$ but grows to $\sim10^{-3}$ during the short initial
transient time before the onset of rapid field growth (see below).
However, for the full non-linear solution we do not observe such a
broad band of instability, modes above $k\simeq 10\pi/L$ are clearly
stable.  (We note also that the spectrum is not very sensitive to the
lattice spacing, which indicates that this cutoff is not a lattice
artifact.) This might be related to the fact that in our simulations
$\theta$ increases only little and $k_{\rm max}$ is not very much
smaller than the particle scale $p_h$. Therefore, it is likely that
back-reaction only allows growth of modes with $k\ll k_{\rm max}$.

\subsection{Chromodynamics}

We now turn to simulations for the non-Abelian $SU(2)$ gauge group.
3d-3v simulations within the hard-loop approximation show that the
exponential growth of transverse magnetic and electric fields
saturates at the scale $m_\infty^2/g$ due to field
self-interactions$\,$\footnote{This, in fact, is true for moderate
anisotropies where the typical longitudinal momentum is less than but
comparable to the typical transverse momentum. For ``extreme''
anisotropies such as~(\ref{AnisoDistrib}) the field strength is
expected to grow larger, see discussion below.}~\cite{Arnold:2005vb,Rebhan:2005re}.
While we can not address such weak fields with PIC methods for
$SU(2)$, we focus here on non-Abelian plasma dynamics beyond the
hard-loop approximation for initial field energy densities of $\sim 10
\, m_\infty^4/g^2$ and above.

\begin{figure}[htb]
\includegraphics[width=3.1in]{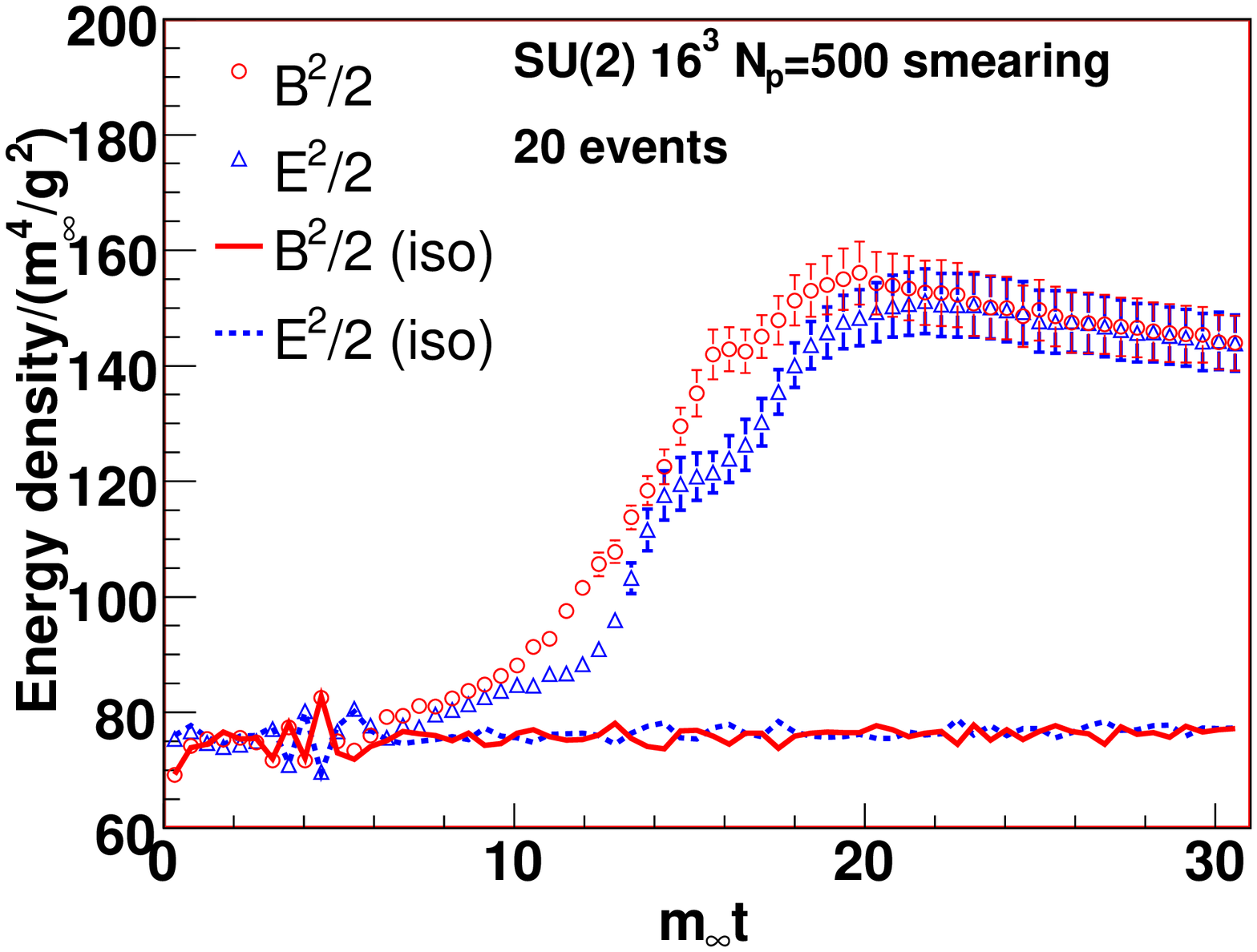}
\includegraphics[width=3.1in]{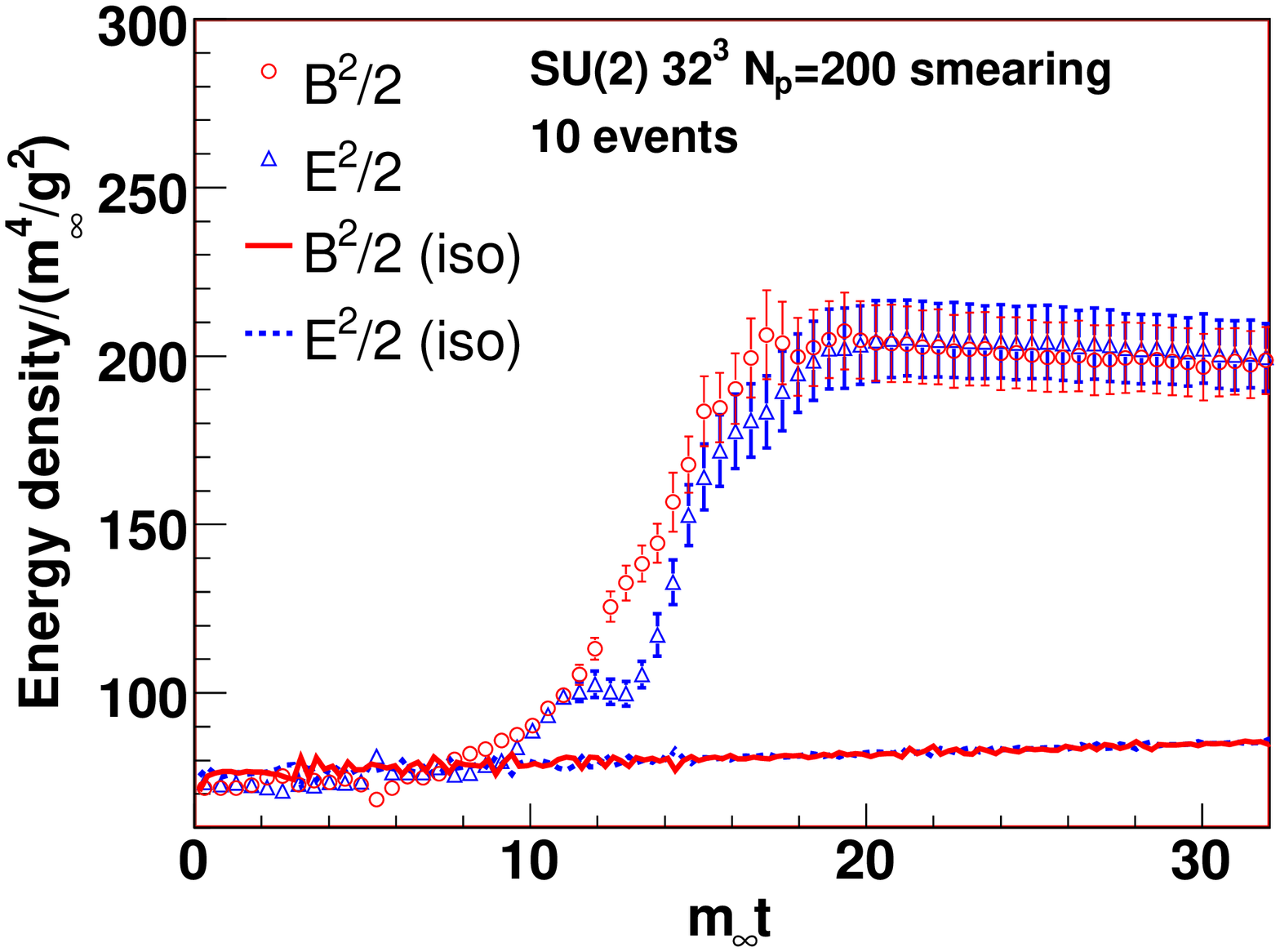}
\includegraphics[width=3.1in]{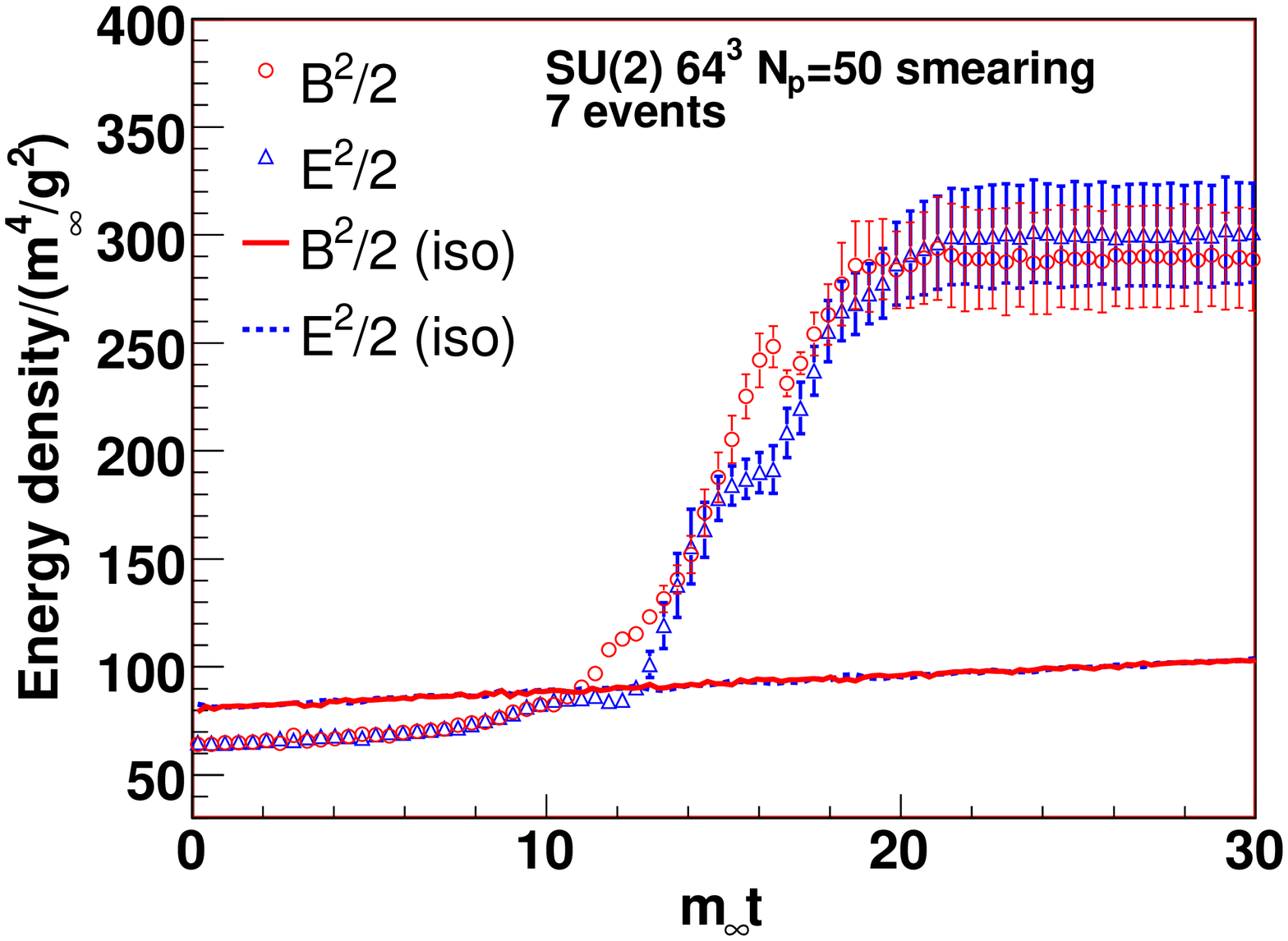}
\includegraphics[width=3.1in]{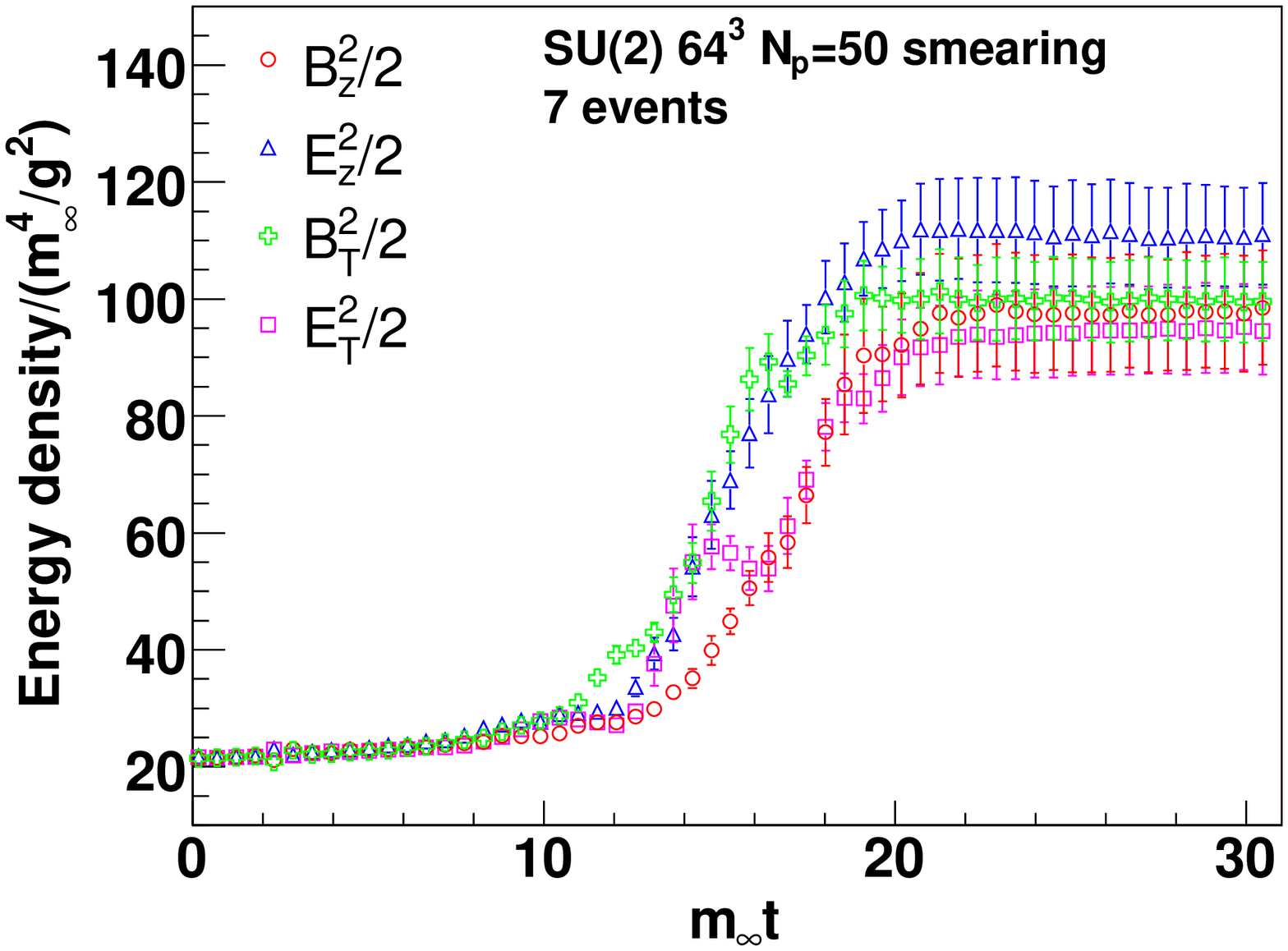}
\caption{Time evolution of the field energy densities for $SU(2)$
gauge group and anisotropic initial particle
momentum distributions.  Simulation parameters are $L=5$~fm, $p_{\rm
hard}=16$~GeV, $g^2\,n_g=10/$fm$^3$, $m_\infty=0.1$~GeV.}
\label{fig:fieldWsu2_weak}
\end{figure}
Fig.~\ref{fig:fieldWsu2_weak} shows results obtained on various
lattices.  In each case the initial condition was taken to be Gaussian
random chromoelectic fields which were ``low-pass'' filtered such that
only the lower half of available lattice modes were populated.
Note that in Fig.~\ref{fig:fieldWsu2_weak} the energy
densities are plotted on a {\em linear} rather than logarithmic scale.
Runs with an isotropic particle distribution are shown in each panel
as an indication of our numerical accuracy (error bars are not
indicated for the isotropic runs).  The isotropic runs show nearly
constant fields over the time interval $t\le 30 \,
m^{-1}_\infty$.\footnote{In the $64^3$ runs the initial amplitudes of
the anisotropic and isotropic energy densities are slightly
different.} From these figures we observe that after a period of rapid
growth both electric and magnetic fields settle to an essentially
constant energy density.

Hence, we confirm that for 3d-3v simulations, even beyond the
hard-loop approximation, the time evolution of non-Abelian fields
stronger than $\sim m_\infty^4/g^2$ differs from that in the
(effectively Abelian) extreme weak-field limit. In particular, a
sustained exponential growth is absent even during the stage where
overall the backreaction on the particles is weak. It should be
emphasized, in fact, that for very strong anisotropies a linear
analysis predicts that the exponential field growth (in the weak-field
situation) could perhaps continue until $B^2\sim m_\infty^4/ g^2/
\Delta\theta^2$, with $\Delta \theta^2 = p_z^2 /
p_T^2$~\cite{Arnold:2005ef}. For our initial
condition~(\ref{AnisoDistrib}), $\Delta\theta^2=0$ at $t=0$ but grows
to ${\cal O}(10^{-3})$ during the initial transient time with constant
fields ($t\,m_\infty\lsim10$ in Fig.~\ref{fig:fieldWsu2_weak}) due to
deflection of the particles.  However, such strong fields are not seen
in our PIC simulations. This may be related to the above-mentioned
back-reaction effects which prevent instability of modes near $k_{\rm
  max} \sim m_\infty/\Delta\theta$.

The results shown in Fig.~\ref{fig:fieldWsu2_weak} indicate a
sensitivity to hard field modes at the ultra-violet end of the
Brioullin zone, $k={\cal O}(a^{-1})$, in contrast to the $U(1)$
simulations shown above and to earlier 1d-3v $SU(2)$
simulations~\cite{Dumitru:2005gp,Dumitru:2005hj,Nara:2005fr}.  The
energy density contained in the fields at late times increases by a
factor of 1.5 when going from a $16^3$ to $32^3$ to $64^3$ lattice
with the same physical size $L$. Hence, the dynamics of $SU(2)$
instabilities seen here is not dominated entirely by a band of
unstable modes in the infrared but clearly involves a cascade of
energy from those modes to a harder scale
$\Lambda$~\cite{Arnold:2005qs}. However, the simulations shown in
Fig.~\ref{fig:fieldWsu2_weak} indicate that $\Lambda(t)$ grows to
${\cal O}(a^{-1})$ during the period of rapid growth of the field
energy density; otherwise, the final field energy density would not
depend on the lattice spacing.

Note that in our PIC simulations the total energy of particles and fields is
conserved, hence $\Lambda(t)$ can not grow arbitrarily large if the
energy density of the fields is UV sensitive. In three spatial
dimensions, this is the case whenever the occupation number of hard
field modes drops more slowly than $\sim1/k^4$. The analysis of
ref.~\cite{Arnold:2005qs}, for example, suggests a $\sim1/k^2$ fall-off
near $k\sim\Lambda$, which indeed corresponds to a quadratically divergent
energy density as $\Lambda\to\infty$. A different
analysis~\cite{Mueller:2006up} suggest a $\sim1/k$ spectrum.

Although energy conservation will eventually stop the growth of the
fields as the lattice spacing decreases towards the continuum, it does
not solve the following problem. When
$\Lambda(t)\sim1/a$, the hard field modes have reached the momentum
scale of the particles, $p_h={\cal O}(a^{-1})$, and so the clean
separation of scales is lost, on which the
approach~(\ref{Vlasov},\ref{YM}) is based. In fact, since the
occupation number (or phase space density) at that scale is of order 1
or less by construction\footnote{If not, the lattice was chosen too
coarse to begin with, and the lattice evolution is far from the continuum
limit.}, it is inappropriate to describe modes at that
scale as a classical field. Those perturbative modes should be
converted dynamically into particles at a lower scale $\Lambda_{f\to
p} \ll 1/a$ so that the field energy density and the entire coupled
field-particle evolution is independent of the artificial lattice
spacing. We hope to be able to study this question in the future.

\begin{figure}[htb]
\includegraphics[width=6.0in]{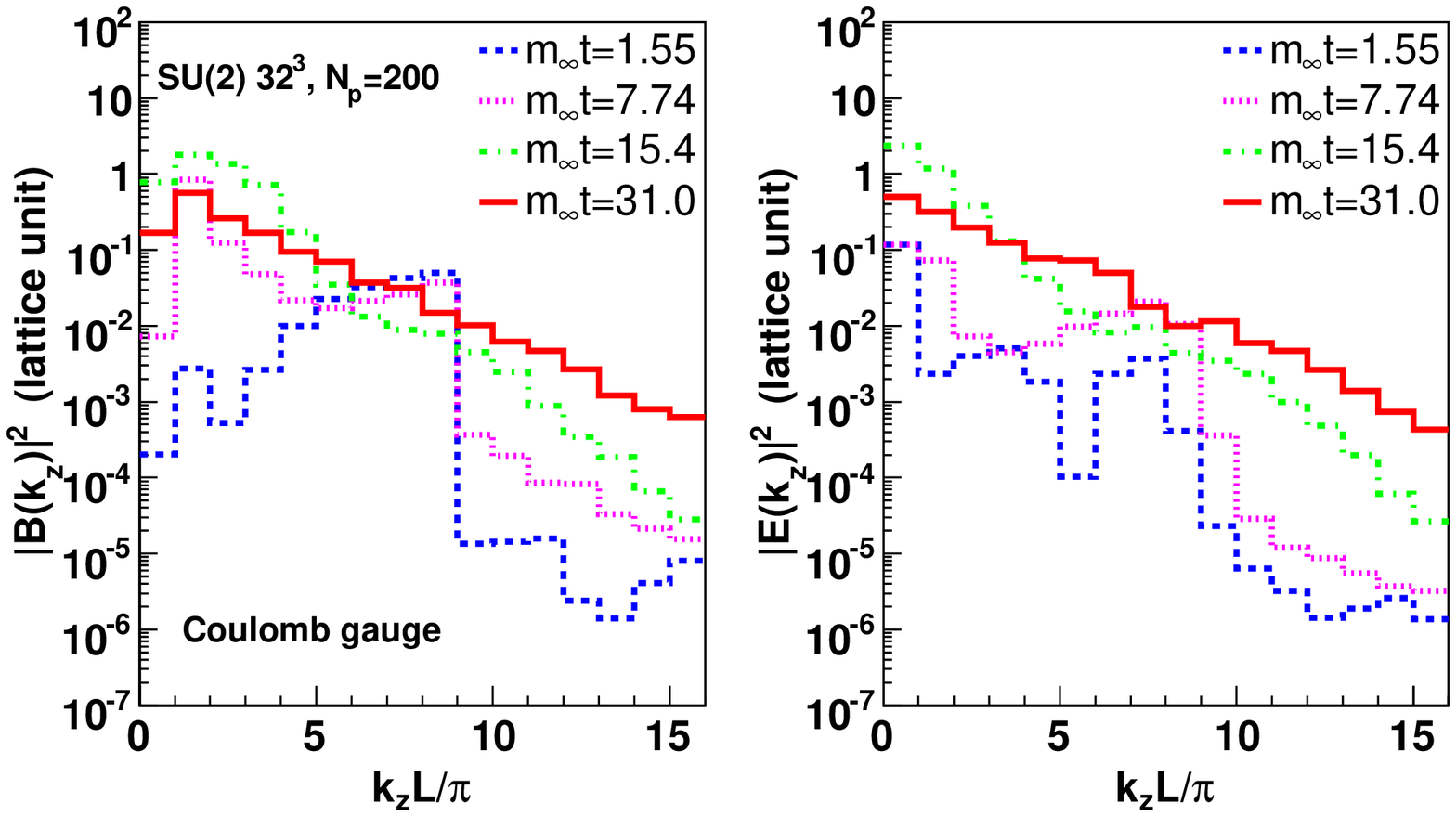}
\includegraphics[width=6.0in]{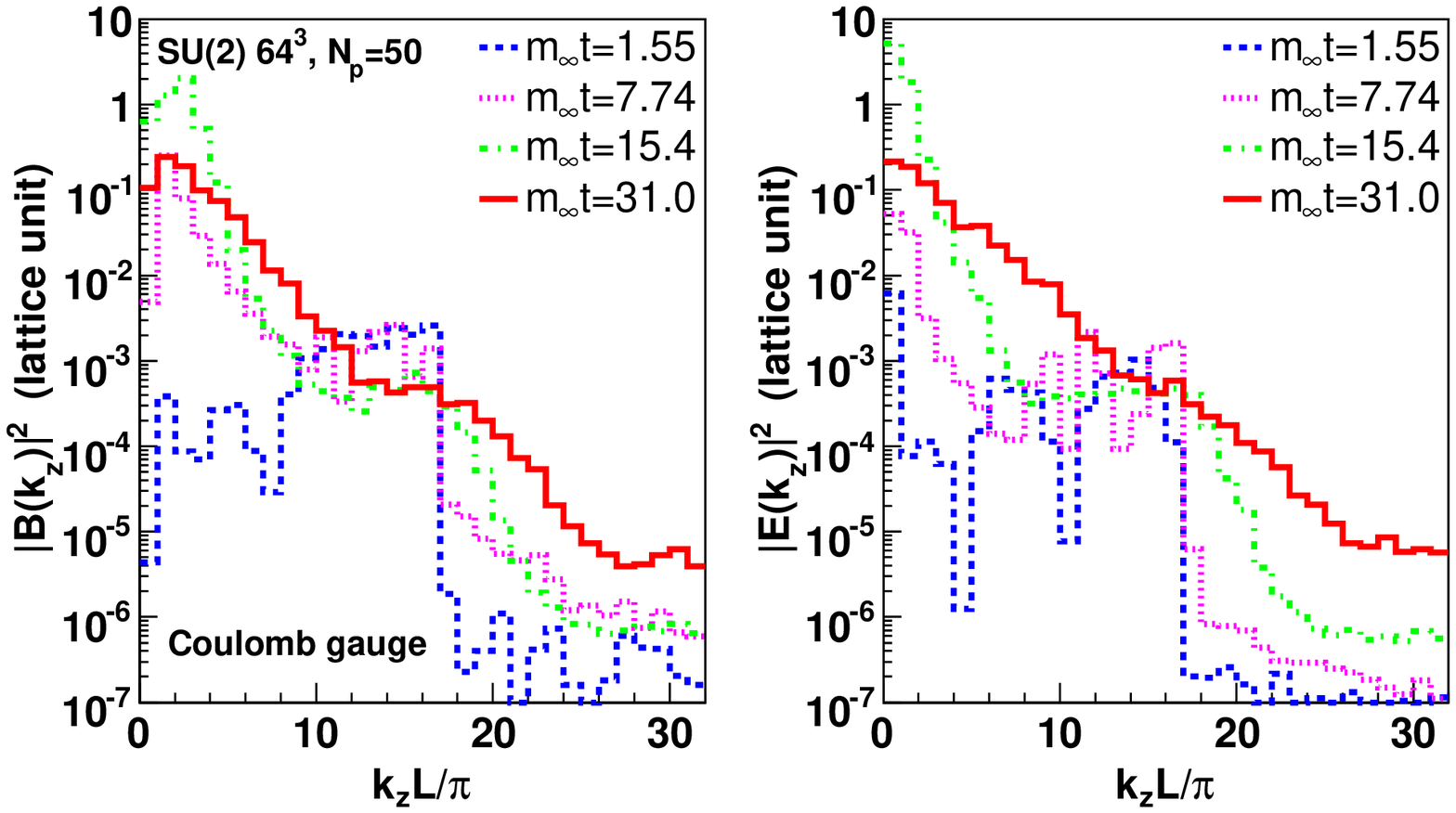}
\caption{Squares of the Fourier transformed (color-) electric and
  magnetic fields in Coulomb gauge (lattice units) at four different times.
  Upper (lower) panels correspond to $32^3$ ($64^3$) lattices. Simulation
  parameters are the same as for Fig.~\ref{fig:fieldWsu2_weak}.
  Note that $\pi/L\approx m_\infty$.}
\label{fig:fieldWsu2_weak_FFT}
\end{figure}
In Fig.~\ref{fig:fieldWsu2_weak} (lower right panel) we have also
plotted the tranverse and longitudinal contributions to the electric
and magnetic energy densities as a function of time.  From this figure
we see that at late times the distribution of field energy obtained
from our 3d-3v $SU(2)$ PIC simulations is approximately isotropic.  In
contrast, for the 1d-3v and 3d-3v $U(1)$ PIC simulations and the 1d-3v
$SU(2)$ CPIC
simulations~\cite{Dumitru:2005gp,Dumitru:2005hj,Nara:2005fr}
transverse magnetic fields dominate, corresponding to an
energy-momentum tensor of the form $T^{00}_{\rm field} \approx
T^{zz}_{\rm field}$ (and $T^{xx}_{\rm field} + T^{yy}_{\rm
field}\approx0$). However, we find that with our 3d-3v $SU(2)$ CPIC
simulations the energy transfered to ``hard'' field modes is
distributed isotropically, regardless of the strong residual
anisotropy of the particle momenta.  The approximate isotropy of the
field energy densities is also seen in HL simulations of strongly
anisotropic plasmas (see below).

In Fig.~\ref{fig:fieldWsu2_weak_FFT} we show
$\bm{E}^a(\bm{k})\cdot\bm{E}^a(-\bm{k})$ and
$\bm{B}^a(\bm{k})\cdot\bm{B}^a(-\bm{k})$ (field spectra) as functions
of $k_z$ at $k_x=k_y=0$. The field configurations at the corresponding
time were gauge transformed to satisfy Coulomb gauge, $\partial_i
A^i=0$ in continuum notation. The Figure shows that before the onset
of the rapid field growth from Fig.~\ref{fig:fieldWsu2_weak} there is
a relatively slow build-up of UV modes, in qualitative agreement with
the discussion in ref.~\cite{Arnold:2005qs}. Also note the strong
growth of the soft (unstable) field modes at early times.  This is the
stage where the field evolution is independent of the lattice
spacing. However, during the period of steep growth from
Fig.~\ref{fig:fieldWsu2_weak} the energy drained from the particles
quickly avalanches from IR to UV field modes and the distribution over
$k$ becomes quite broad. We attribute this avalanche to the UV to the
self-interaction of the Yang-Mills fields since it is not observed for
Abelian $U(1)$ fields (Fig.~\ref{fig:FT_fieldWu1_weak}) even for
extreme particle anisotropies.

\begin{figure}[htb]
\includegraphics[width=4.0in]{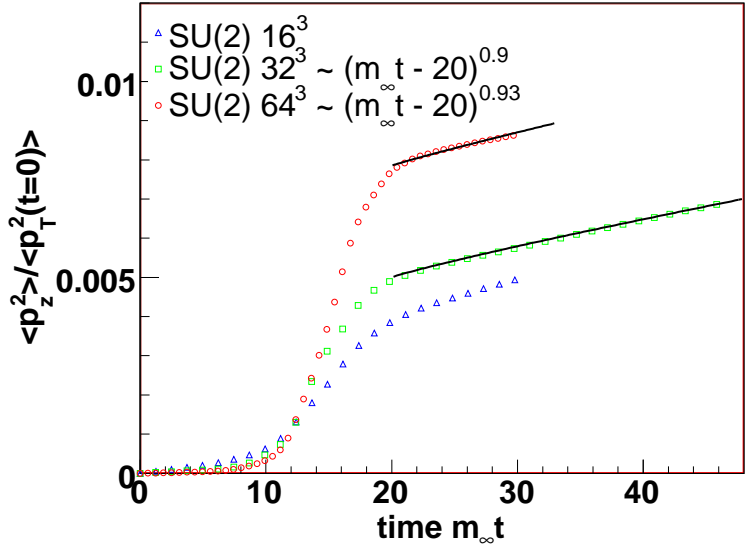}
\caption{Broadening of the longitudinal particle
momentum distribution for the simulations from
Fig.~\ref{fig:fieldWsu2_weak}. The line represents the power-law fit
from eq.~(\ref{eq:theta-fit}).}
\label{fig:thetaSU2_weak}
\end{figure}
Fig.~\ref{fig:thetaSU2_weak} depicts the time evolution of the typical
longitudinal particle momentum. The broadening of the longitudinal
distribution during the initial transient $t\,m_\infty\lsim10$ is {\em
independent} of the lattice spacing. During the stage of steep
field growth, $p_z$ broadens more rapidly, especially for the
finer $64^3$ lattice. Since
\begin{equation}
\langle p_z^2\rangle \sim t\, \frac{dN_{scatt}}{dt}\,q_z^2~,
\end{equation}
with $q_z$ the typical momentum transfer in $z$-direction and
${dN_{scatt}}/{dt}$ the collision rate of particles with the field
modes, it follows that during this stage either the scattering rate
and/or the typical momentum transfer increase. This observation (and
the fact that the effect is much stronger for the $64^3$ lattice) is
consistent with the idea that during this stage the UV-cutoff
$\Lambda(t)$ for the classical field grows to ${\cal O}(a^{-1})$.

A power-law fit to the late-time evolution of
$\langle p_z^2 \rangle$ of the form
\begin{equation}  \label{eq:theta-fit}
{\frac{\langle p_z^2\rangle}{\langle p_x^2+p_y^2\rangle}} 
   \propto t^\alpha~,
\end{equation}
yields $\alpha\simeq1$, corresponding to a random walk of the
particles in $p_z$ with constant collision rate and average momentum
transfer.\footnote{Hence, $\Lambda$ appears to be nearly constant in
the late stages, which is natural if it has already grown to ${\cal
O}(a^{-1})$.} In an expanding metric, this should then lead to a
$\sim\tau^{-1/4}$ drop of the typical $p_z$, as argued by
B\"odeker~\cite{Bodeker:2005nv} (see also section~V in~\cite{Arnold:2005ef}).  Hence,
the anisotropy of the hard modes which develops in the early stage of
a heavy-ion collision due to the rapid longitudinal expansion should
be somewhat smaller than expected within the original ``bottom-up''
scenario~\cite{Baier:2000sb}, which predicted $\surd\langle p_z^2 \rangle
\sim \tau^{-1/3}$.

\begin{figure}[htb]
\includegraphics[width=3.0in]{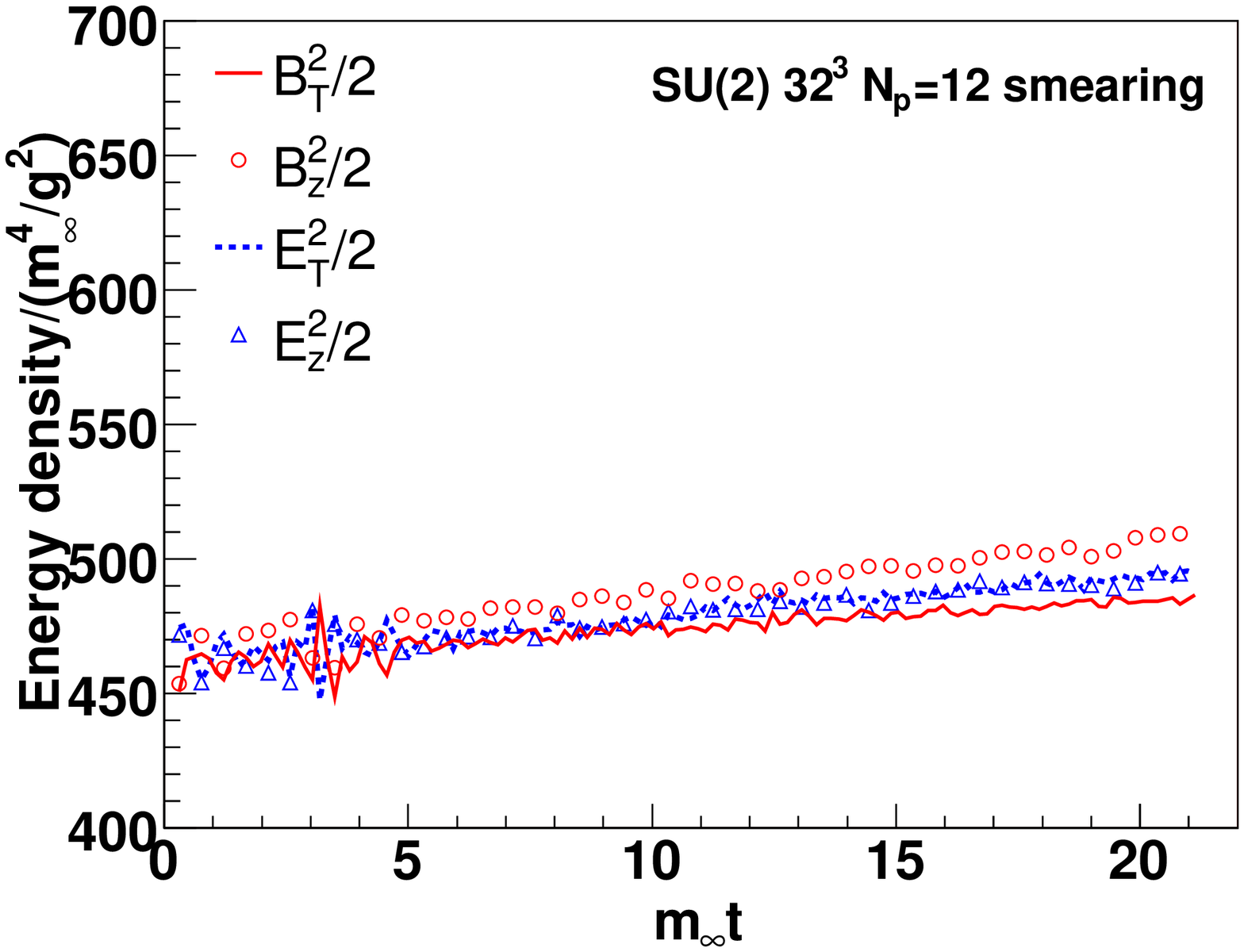}
\includegraphics[width=3.0in]{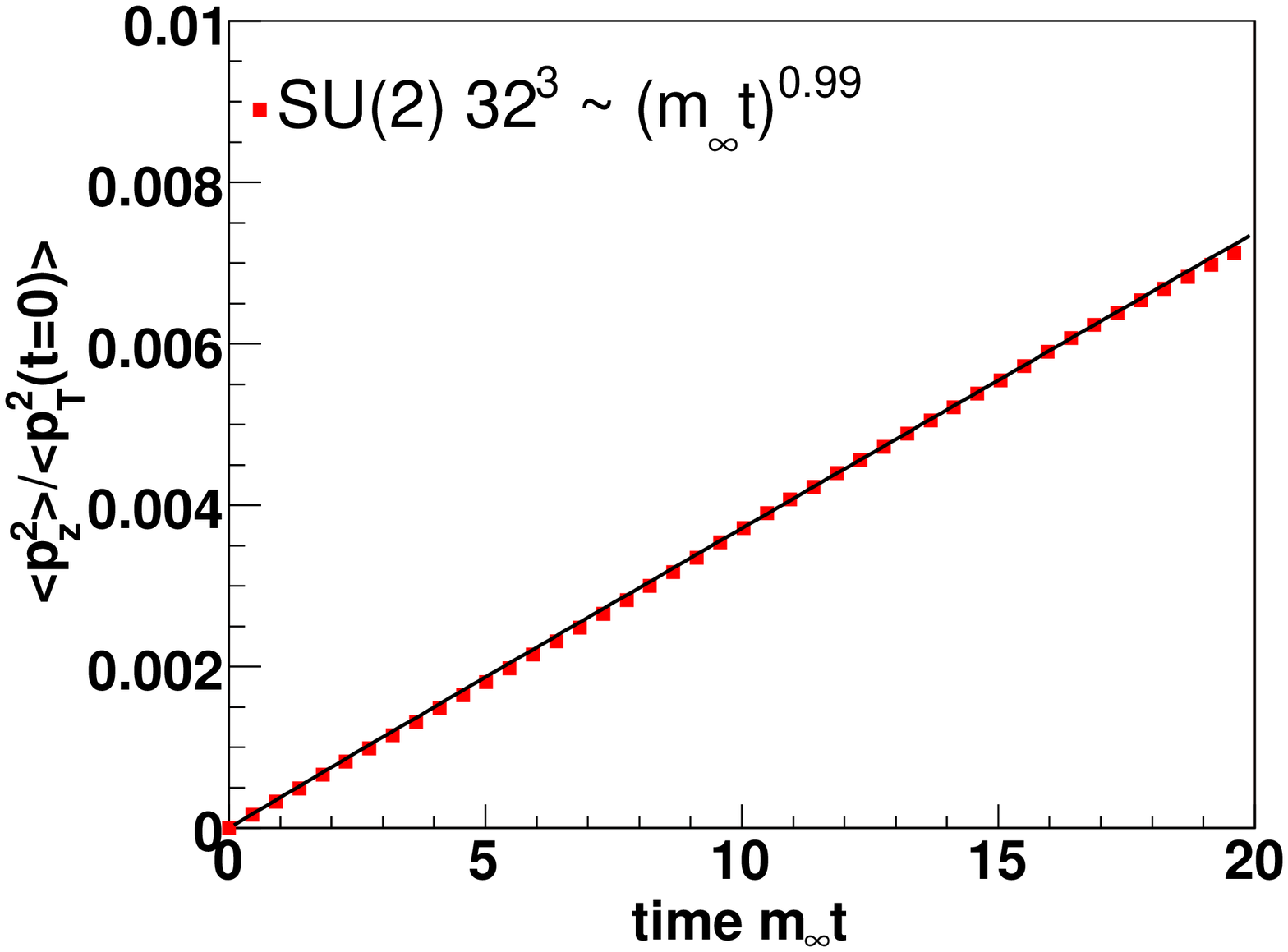}
\caption{Left: field evolution for stronger initial fields.
Right: evolution of the average longitudinal particle momentum.
The line represents the power-law
fit from eq.~(\ref{eq:theta-fit}). The remaining simulation parameters are
$p_h=16$~GeV, $g^2\,n_g=10$/fm$^3$, $L=5$~fm, as before.}
\label{fig:fieldWsu2_strong}
\end{figure}
For even stronger initial fields, we find that the growth rate of the
electric and magnetic fields and the dependence on the lattice spacing 
diminishes.
Only a very mild linear growth remains over time spans which we can
access in our simulations, see Fig.~\ref{fig:fieldWsu2_strong}.  A
power-law fit of the form~(\ref{eq:theta-fit}) again shows that the
variance of the longitudinal momentum distribution grows linearly in
time, corresponding to a random walk of the particles in $p_z$ with
constant scattering rate (off the field modes) and typical momentum
transfer.

\begin{figure}[htb]
\includegraphics[width=4.0in]{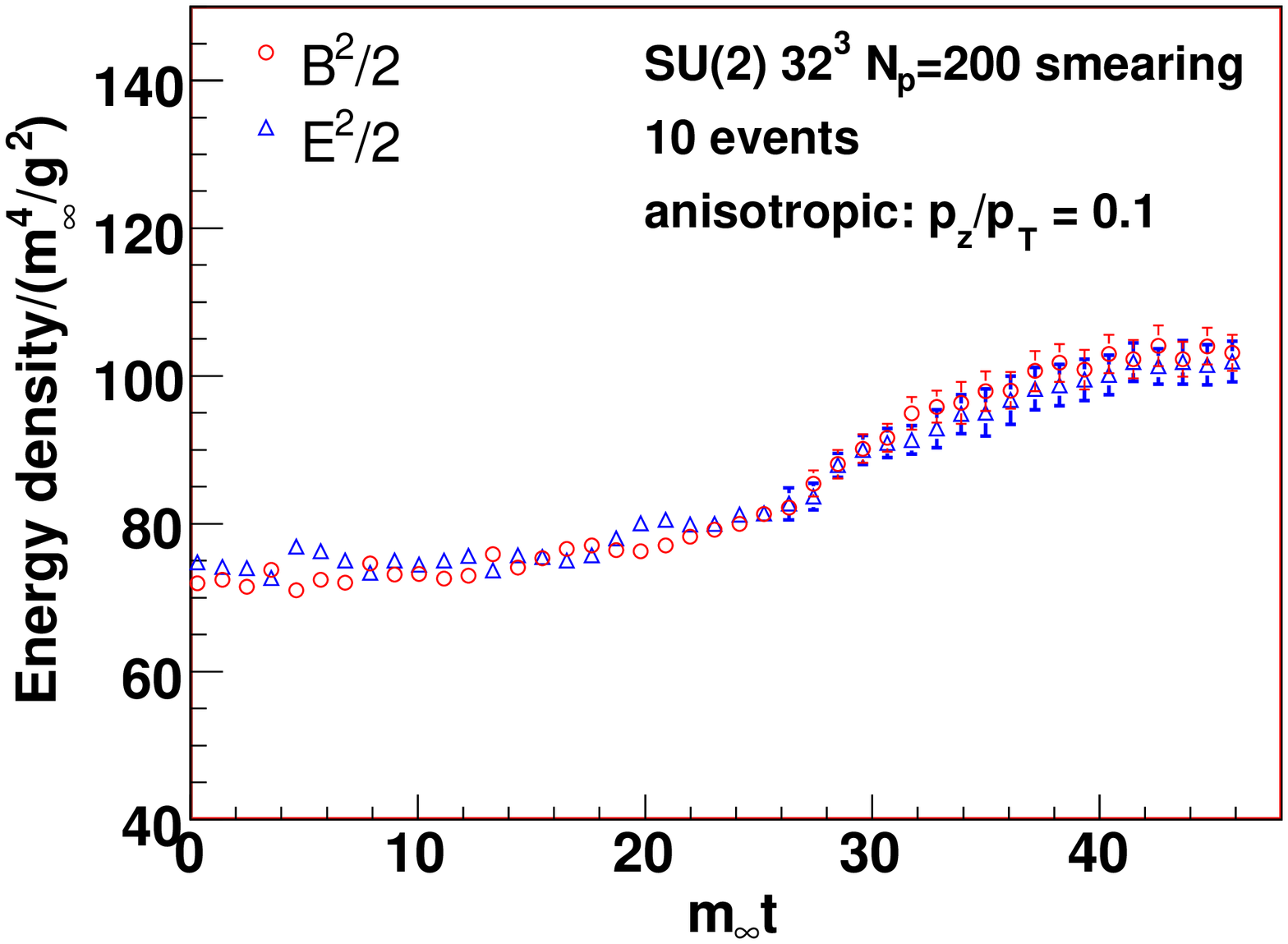}
\includegraphics[width=5.0in]{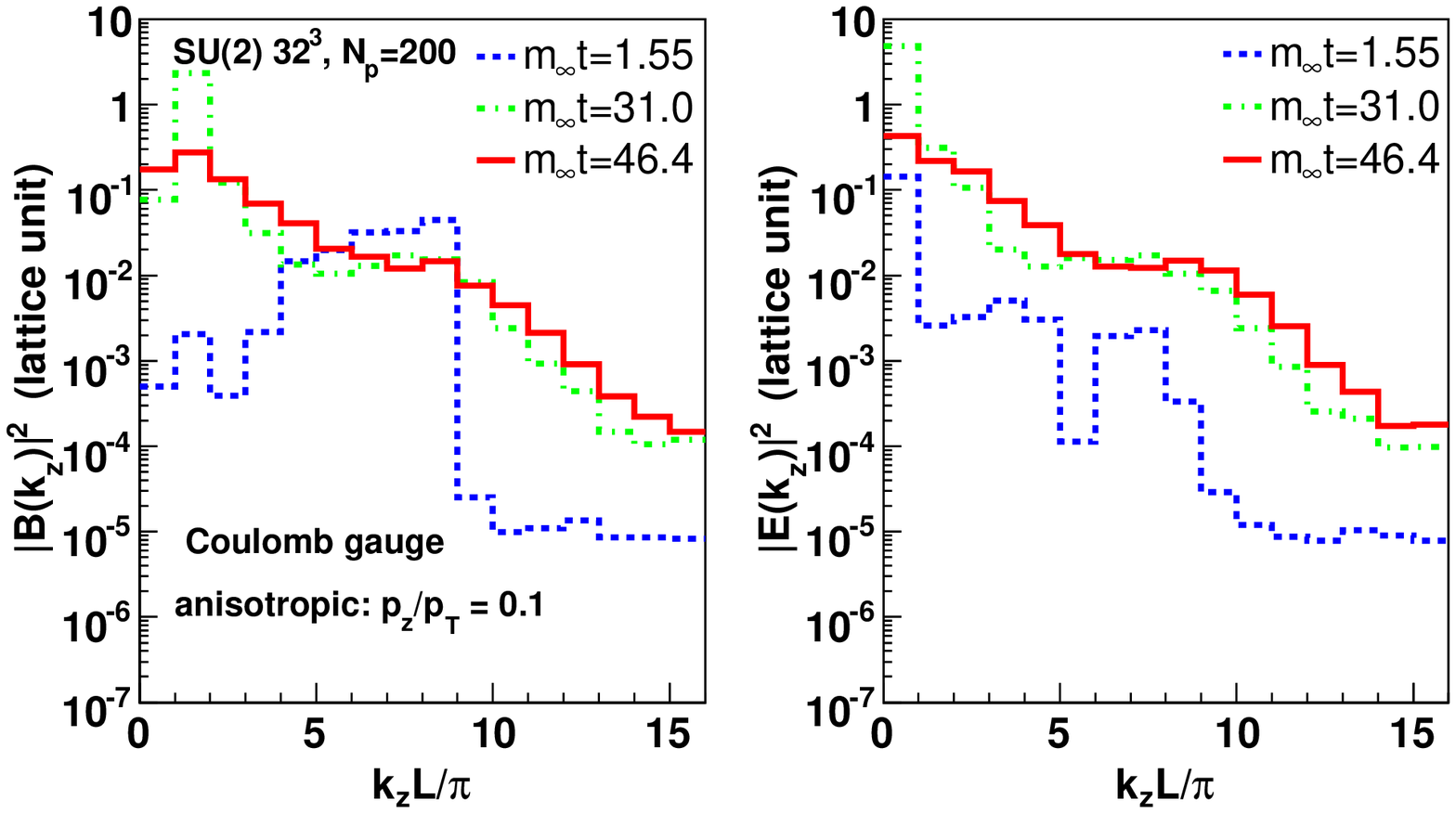}
\caption{Field evolution for moderate initial anisotropy,
$\sqrt{\langle p_z^2\rangle/\langle p_T^2\rangle}=10\%$.  The
remaining simulation parameters are the same as in
Fig.~\ref{fig:fieldWsu2_weak}. Top: field energy density as a function
of time.  Bottom: Fourier transformed color-fields (in Coulomb gauge)
at different times.}
\label{fig:moderateAnisotropy}
\end{figure}
We have also performed a simulation with a more moderate initial
anisotropy of $\sqrt{\langle p_z^2\rangle/\langle p_T^2\rangle}=10\%$,
replacing the function $\delta(p_z)$ in~(\ref{AnisoDistrib}) by an
exponential $\sim\exp(-10\,|p_z|/p_h)$.  The time evolution of the
fields and of their Fourier transform is shown in
Fig.~\ref{fig:moderateAnisotropy}.  As expected, the onset of field
growth is delayed, by roughly a factor of two, as compared to the
simulation with extreme particle anisotropy. Also, the intermediate
rapid growth of the fields is weakened since the rate of energy
transfer from particles to fields is now lower. Nevertheless, even for
this run with more moderate anisotropy, we observe that the Fourier
spectrum flattens somewhat at late times (when the growth saturates);
it actually resembles the final spectrum from
Fig.~\ref{fig:fieldWsu2_weak_FFT} although the angular width of the
particle momentum distribution is now much larger.

\subsubsection*{Comparison with hard-loop results}

\begin{figure}[htb]
\includegraphics[width=4in]{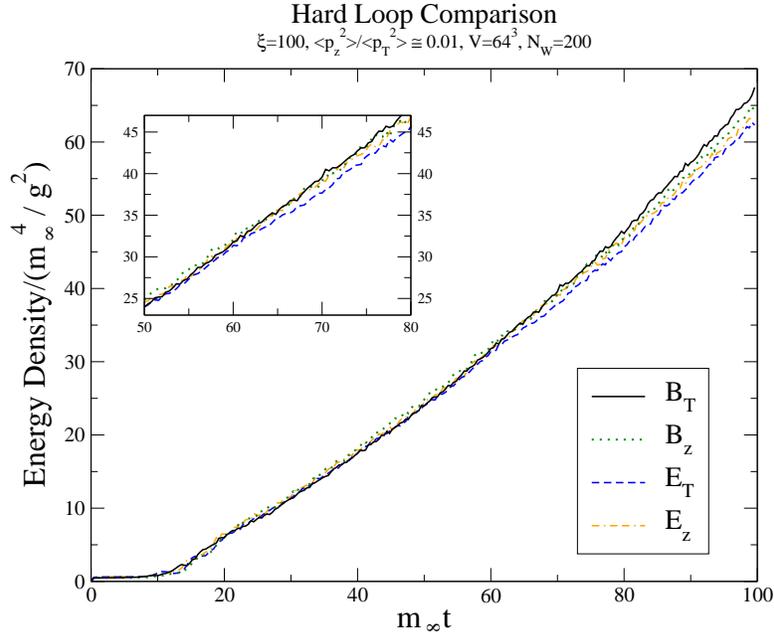}
\caption{Field evolution for moderate initial anisotropy as obtained
from numerical solution of hard-loop effective theory
\cite{Rebhan:2005re}.}
\label{fig:hlCompare}
\end{figure}
In Fig.~\ref{fig:hlCompare} we present the dependence of the energy 
densities obtained by solving the soft-field equations of motion in 
the hard-loop approximation \cite{Rebhan:2005re} for the case of 
$\sqrt{\langle p_z^2 \rangle / \langle p_T^2 \rangle} \sim 0.1$.  For 
this simulation the initial condition has no fields, but instead has 
finite fluctuations of the auxilliary $W$ fields which induce field 
fluctuations with typical initial energy density ${\cal E} \sim 0.5\, 
m_\infty^4/g^2$ at early times (for details of the initial condition 
and numerical algorithm used see Ref.~\cite{Rebhan:2005re}).  Note 
that if one makes the same run with an isotropic hard particle 
distribution the fields remain isotropic and show no signs of growth.

As can be seen from Fig.~\ref{fig:hlCompare} there is an initial
period where no field growth occurs, similar to
Fig.~\ref{fig:moderateAnisotropy}. At $m_\infty t \sim 10$ all field
components begin to grow isotropically in the hard-loop simulation.
They then grow approximately linearly and isotropically for the
duration of the run.  At late times one can see a slight curvature
upwards suggesting a faster than linear growth.  This curvature is
related to the finite discretizations used in both space-time and
velocity space.  Improving these discretizations or damping high
angular momentum modes~\cite{Arnold:2005vb} extends the region in time
over which the linear behavior is observed.  The inset of
Fig.~\ref{fig:hlCompare} shows the dependence of the field energy
densities at late times when the magnitudes are comparable to those
shown in Fig.~\ref{fig:moderateAnisotropy}.\footnote{Note that to
match to the total energy density take into account the factor of
three difference between the total and the contributions from
individual components.}  Also, similar to the CPIC results we see that
the energy densities obtained in the hard-loop simulation are
approximately isotropic.

\section{Discussion}
\label{sec_Discussion}

We have presented numerical solutions of coupled particle-field
dynamics in Abelian and non-Abelian gauge theories beyond the HL
approximation. Such plasmas exhibit collective instabilities when the
distribution of particles in momentum space is anisotropic (we
restrict ourselves to an oblate distribution here). However, in the
regime where field self-interactions can not be neglected,
corresponding to field strengths of order $\sim m_\infty^2/g$ and
above, the evolution in chromodynamics is rather different from that
in Abelian plasmas. Most remarkably, for the initial conditions
employed here, we do not observe a period of sustained exponential
growth of the soft fields. This is because the field modes can
interact directly and so the energy drained by the instability from
the hard particles does not remain in unstable modes for a
sufficiently long time. 

The transition from exponential to linear growth at the scale
$m_\infty^2/g$ has been seen previously in 3d-3v simulations within
the HL approximation~\cite{Arnold:2005vb,Rebhan:2005re}. It appears
that if the particles exhibit only weak anisotropy, that the HL
approach could be continued until very late times since hard field
modes are populated very slowly. However, for moderate anisotropies
the instability is also affected more easily by collisions among the
particles~\cite{Schenke:2006xu} and by the rapid longitudinal
expansion encountered in relativistic heavy-ion
collisions~\cite{Romatschke:2006wg}.

For very anisotropic particle momentum distributions we find
here that the evolution differs significantly for field strengths of
order $m_\infty^2/g$ and above, even if overall the back-reaction on
the hard particle modes is small. After some initial transient time,
we observe a short period of rapid growth of soft electric and
magnetic fields followed by an avalanche of energy into the UV. As
long as the energy of the classical Yang-Mills fields is still much
less than that of the particles, this avalanche proceeds to the
smallest length scales, namely the lattice spacing $a$, before
reaching a steady-state distribution. Since $1/a$ will eventually
reach $p_h$ as one takes the continuum limit, the loss of a clear
separation of scales between soft and hard degrees of freedom could
perhaps be viewed as (partial) ``classical decay'' of the field into
particles. To understand what happens then requires us to go beyond
the Wong-Yang-Mills equations, which rely on such a separation of
scales. We also point out that the energy-momentum tensor of the
fields is nearly isotropic, while Weibel instabilities seen in $U(1)$
and 1d-3v $SU(2)$ simulations lead to dominance of transverse magnetic
fields.

If the initial field energy density is rather high, of order
$(10^2-10^3)\,m^4_\infty/g^2$, we do not observe an instability with
rapid growth of the fields. In this case, the mean-square width of the
longitudinal particle distribution grows linearly in time. This
corresponds to a random walk with constant scattering rate (from the
field fluctuations) and typical momentum transfer. For even stronger
initial fields with energy density comparable to that of the
particles, $\sim p_h\,n_g$, we of course expect a very rapid
isotropization of the particle pressure, see e.g.\
refs.~\cite{Dumitru:2005gp,Dumitru:2005hj,Nara:2005fr}.

Our simulations indicate that it could be important to implement a
dynamical field to particle conversion at a scale $\Lambda_{f\to p}
< p_h$ to avoid that field and particle degrees of freedom overlap. This
might lead to a continuous ``cycle'' of energy transfer from the
anisotropic particles to soft field modes which then quickly
re-emerges at that scale in isotropic form. The cycle should
continue until all particle momenta are isotropic. Theoretically, it
also remains to be understood why the energy avalanche from soft to
hard field modes occurs so rapidly. An independent analytical
discussion of energy transfer to small length scales in QCD has been
recently given in ref.~\cite{Mueller:2006up}.  In the
future, we intend to perform more quantitative studies and verify the
matching to the ``bottom-up'' solution for particle-field
thermalization~\cite{Mueller:2005hj}. To do so, however, requires us
to also allow for longitudinal expansion of the lattice and to include
hard elastic and inelastic scattering of particles.

\section*{Acknowledgements}
M.S.\ thanks A.~Rebhan and P.~Romatschke for
their collaboration. A.D.\ and Y.N.\ thank Larry McLerran for
discussions and continuous encouragement to study non-perturbative
particle-gauge field systems. Y.N.\ acknowledges support from GSI and DFG.
Numerical computations were performed at the Center for Scientific
Computing (CSC) at Frankfurt University.

\appendix

\section{Momentum update} 
\label{appendixMomUpdate}

We consider the discretization of the equation
\begin{equation}
  \frac{d\bm{p}}{dt} = gQ^a(\bm{E}^a + \bm{v}\times \bm{B}^a).
\end{equation}
The difference form of the equation is
\begin{equation}
  \frac{\bm{p}(t+\Delta t/2)-\bm{p}(t-\Delta t/2)}{\Delta t} = 
  gQ^a\left(\bm{E}^a(t)
    + \frac{\bm{p}(t+\Delta t/2)+\bm{p}(t-\Delta t/2)}{2}
     \times \frac{\bm{B}^a(t)}{e(t)}  \right) \label{eq:mom}
\end{equation}
Eq.(\ref{eq:mom}) can be solved by the Buneman-Boris method~\cite{Hockney:1985,Birdsall:1985}
as follows:
\begin{eqnarray}
  \bm{p}(t) &=& \bm{p}(t-\Delta t/2) + \frac{\Delta t}{2}\bm{E}(t),\\
  e(t) &=& |\bm{p}(t)|,\\
  \bm{p}'(t) &=& \bm{p}(t) + \frac{\Delta t}{2}
                 \bm{p}(t)\times\frac{\bm{B}(t)}{e(t)},\\
  \bm{p}_2(t) &=& \bm{p}(t) + \frac{2}{1+(\bm{B}/e(t)\Delta t/2)^2}
          \frac{\Delta t}{2}\bm{p}'(t)
                 \times\frac{\bm{B}(t)}{e(t)},\\
  \bm{p}(t+\Delta t/2) &=& \bm{p}_2(t) + \frac{\Delta t}{2}\bm{E}(t),
\end{eqnarray}
where $\bm{E}\equiv gQ^a\bm{E}^a$ and $\bm{B}\equiv gQ^a\bm{B}^a$.
This scheme is time reversible and the overall momentum
integration is accurate to second-order in the time step.


\bibliography{ms_3dWYM}

\begin{thebibliography}{55}
\expandafter\ifx\csname natexlab\endcsname\relax\def\natexlab#1{#1}\fi
\expandafter\ifx\csname bibnamefont\endcsname\relax
  \def\bibnamefont#1{#1}\fi
\expandafter\ifx\csname bibfnamefont\endcsname\relax
  \def\bibfnamefont#1{#1}\fi
\expandafter\ifx\csname citenamefont\endcsname\relax
  \def\citenamefont#1{#1}\fi
\expandafter\ifx\csname url\endcsname\relax
  \def\url#1{\texttt{#1}}\fi
\expandafter\ifx\csname urlprefix\endcsname\relax\def\urlprefix{URL }\fi
\providecommand{\bibinfo}[2]{#2}
\providecommand{\eprint}[2][]{\url{#2}}

\bibitem[{\citenamefont{Mueller}(2003)}]{Mueller:2002kw}
\bibinfo{author}{\bibfnamefont{A.~H.} \bibnamefont{Mueller}},
  \bibinfo{journal}{Nucl. Phys.} \textbf{\bibinfo{volume}{A715}},
  \bibinfo{pages}{20} (\bibinfo{year}{2003}), \eprint{hep-ph/0208278}.

\bibitem[{\citenamefont{Iancu and Venugopalan}(2003)}]{Iancu:2003xm}
\bibinfo{author}{\bibfnamefont{E.}~\bibnamefont{Iancu}} \bibnamefont{and}
  \bibinfo{author}{\bibfnamefont{R.}~\bibnamefont{Venugopalan}}
  (\bibinfo{year}{2003}), \eprint{hep-ph/0303204}.

\bibitem[{\citenamefont{McLerran}(2005)}]{McLerran:2005kk}
\bibinfo{author}{\bibfnamefont{L.}~\bibnamefont{McLerran}},
  \bibinfo{journal}{Nucl. Phys.} \textbf{\bibinfo{volume}{A752}},
  \bibinfo{pages}{355} (\bibinfo{year}{2005}).

\bibitem[{\citenamefont{Baier et~al.}(2001)\citenamefont{Baier, Mueller,
  Schiff, and Son}}]{Baier:2000sb}
\bibinfo{author}{\bibfnamefont{R.}~\bibnamefont{Baier}},
  \bibinfo{author}{\bibfnamefont{A.~H.} \bibnamefont{Mueller}},
  \bibinfo{author}{\bibfnamefont{D.}~\bibnamefont{Schiff}}, \bibnamefont{and}
  \bibinfo{author}{\bibfnamefont{D.~T.} \bibnamefont{Son}},
  \bibinfo{journal}{Phys. Lett.} \textbf{\bibinfo{volume}{B502}},
  \bibinfo{pages}{51} (\bibinfo{year}{2001}), \eprint{hep-ph/0009237}.

\bibitem[{\citenamefont{Mr{\'o}wczy{\'n}ski}(1993)}]{Mrowczynski:1993qm}
\bibinfo{author}{\bibfnamefont{S.}~\bibnamefont{Mr{\'o}wczy{\'n}ski}},
  \bibinfo{journal}{Phys. Lett.} \textbf{\bibinfo{volume}{B314}},
  \bibinfo{pages}{118} (\bibinfo{year}{1993}).

\bibitem[{\citenamefont{Mr{\'o}wczy{\'n}ski}(1994)}]{Mrowczynski:1994xv}
\bibinfo{author}{\bibfnamefont{S.}~\bibnamefont{Mr{\'o}wczy{\'n}ski}},
  \bibinfo{journal}{Phys. Rev.} \textbf{\bibinfo{volume}{C49}},
  \bibinfo{pages}{2191} (\bibinfo{year}{1994}).

\bibitem[{\citenamefont{Mr{\'o}wczy{\'n}ski}(1997)}]{Mrowczynski:1996vh}
\bibinfo{author}{\bibfnamefont{S.}~\bibnamefont{Mr{\'o}wczy{\'n}ski}},
  \bibinfo{journal}{Phys. Lett.} \textbf{\bibinfo{volume}{B393}},
  \bibinfo{pages}{26} (\bibinfo{year}{1997}), \eprint{hep-ph/9606442}.

\bibitem[{\citenamefont{Romatschke and Strickland}(2003)}]{Romatschke:2003ms}
\bibinfo{author}{\bibfnamefont{P.}~\bibnamefont{Romatschke}} \bibnamefont{and}
  \bibinfo{author}{\bibfnamefont{M.}~\bibnamefont{Strickland}},
  \bibinfo{journal}{Phys. Rev.} \textbf{\bibinfo{volume}{D68}},
  \bibinfo{pages}{036004} (\bibinfo{year}{2003}), \eprint{hep-ph/0304092}.

\bibitem[{\citenamefont{Arnold et~al.}(2003)\citenamefont{Arnold, Lenaghan, and
  Moore}}]{Arnold:2003rq}
\bibinfo{author}{\bibfnamefont{P.}~\bibnamefont{Arnold}},
  \bibinfo{author}{\bibfnamefont{J.}~\bibnamefont{Lenaghan}}, \bibnamefont{and}
  \bibinfo{author}{\bibfnamefont{G.~D.} \bibnamefont{Moore}},
  \bibinfo{journal}{JHEP} \textbf{\bibinfo{volume}{08}}, \bibinfo{pages}{002}
  (\bibinfo{year}{2003}), \eprint{hep-ph/0307325}.

\bibitem[{\citenamefont{Romatschke and Strickland}(2004)}]{Romatschke:2004jh}
\bibinfo{author}{\bibfnamefont{P.}~\bibnamefont{Romatschke}} \bibnamefont{and}
  \bibinfo{author}{\bibfnamefont{M.}~\bibnamefont{Strickland}},
  \bibinfo{journal}{Phys. Rev.} \textbf{\bibinfo{volume}{D70}},
  \bibinfo{pages}{116006} (\bibinfo{year}{2004}), \eprint{hep-ph/0406188}.

\bibitem[{\citenamefont{Mr{\'o}wczy{\'n}ski
  et~al.}(2004)\citenamefont{Mr{\'o}wczy{\'n}ski, Rebhan, and
  Strickland}}]{Mrowczynski:2004kv}
\bibinfo{author}{\bibfnamefont{S.}~\bibnamefont{Mr{\'o}wczy{\'n}ski}},
  \bibinfo{author}{\bibfnamefont{A.}~\bibnamefont{Rebhan}}, \bibnamefont{and}
  \bibinfo{author}{\bibfnamefont{M.}~\bibnamefont{Strickland}},
  \bibinfo{journal}{Phys. Rev.} \textbf{\bibinfo{volume}{D70}},
  \bibinfo{pages}{025004} (\bibinfo{year}{2004}), \eprint{hep-ph/0403256}.

\bibitem[{\citenamefont{Romatschke and Venugopalan}(2005)}]{Romatschke:2005ag}
\bibinfo{author}{\bibfnamefont{P.}~\bibnamefont{Romatschke}} \bibnamefont{and}
  \bibinfo{author}{\bibfnamefont{R.}~\bibnamefont{Venugopalan}}
  (\bibinfo{year}{2005}), \eprint{hep-ph/0510292}.

\bibitem[{\citenamefont{Romatschke and
  Venugopalan}(2006{\natexlab{a}})}]{Romatschke:2005pm}
\bibinfo{author}{\bibfnamefont{P.}~\bibnamefont{Romatschke}} \bibnamefont{and}
  \bibinfo{author}{\bibfnamefont{R.}~\bibnamefont{Venugopalan}},
  \bibinfo{journal}{Phys. Rev. Lett.} \textbf{\bibinfo{volume}{96}},
  \bibinfo{pages}{062302} (\bibinfo{year}{2006}{\natexlab{a}}),
  \eprint{hep-ph/0510121}.

\bibitem[{\citenamefont{Romatschke and
  Venugopalan}(2006{\natexlab{b}})}]{Romatschke:2006nk}
\bibinfo{author}{\bibfnamefont{P.}~\bibnamefont{Romatschke}} \bibnamefont{and}
  \bibinfo{author}{\bibfnamefont{R.}~\bibnamefont{Venugopalan}}
  (\bibinfo{year}{2006}{\natexlab{b}}), \eprint{hep-ph/0605045}.

\bibitem[{\citenamefont{Arnold et~al.}(2005)\citenamefont{Arnold, Moore, and
  Yaffe}}]{Arnold:2005vb}
\bibinfo{author}{\bibfnamefont{P.}~\bibnamefont{Arnold}},
  \bibinfo{author}{\bibfnamefont{G.~D.} \bibnamefont{Moore}}, \bibnamefont{and}
  \bibinfo{author}{\bibfnamefont{L.~G.} \bibnamefont{Yaffe}},
  \bibinfo{journal}{Phys. Rev.} \textbf{\bibinfo{volume}{D72}},
  \bibinfo{pages}{054003} (\bibinfo{year}{2005}), \eprint{hep-ph/0505212}.

\bibitem[{\citenamefont{Rebhan et~al.}(2005)\citenamefont{Rebhan, Romatschke,
  and Strickland}}]{Rebhan:2005re}
\bibinfo{author}{\bibfnamefont{A.}~\bibnamefont{Rebhan}},
  \bibinfo{author}{\bibfnamefont{P.}~\bibnamefont{Romatschke}},
  \bibnamefont{and}
  \bibinfo{author}{\bibfnamefont{M.}~\bibnamefont{Strickland}},
  \bibinfo{journal}{JHEP} \textbf{\bibinfo{volume}{09}}, \bibinfo{pages}{041}
  (\bibinfo{year}{2005}), \eprint{hep-ph/0505261}.

\bibitem[{\citenamefont{Romatschke and Rebhan}(2006)}]{Romatschke:2006wg}
\bibinfo{author}{\bibfnamefont{P.}~\bibnamefont{Romatschke}} \bibnamefont{and}
  \bibinfo{author}{\bibfnamefont{A.}~\bibnamefont{Rebhan}}
  (\bibinfo{year}{2006}), \eprint{hep-ph/0605064}.

\bibitem[{\citenamefont{Krasnitz and Venugopalan}(1999)}]{Krasnitz:1998ns}
\bibinfo{author}{\bibfnamefont{A.}~\bibnamefont{Krasnitz}} \bibnamefont{and}
  \bibinfo{author}{\bibfnamefont{R.}~\bibnamefont{Venugopalan}},
  \bibinfo{journal}{Nucl. Phys.} \textbf{\bibinfo{volume}{B557}},
  \bibinfo{pages}{237} (\bibinfo{year}{1999}), \eprint{hep-ph/9809433}.

\bibitem[{\citenamefont{Krasnitz and Venugopalan}(2000)}]{Krasnitz:1999wc}
\bibinfo{author}{\bibfnamefont{A.}~\bibnamefont{Krasnitz}} \bibnamefont{and}
  \bibinfo{author}{\bibfnamefont{R.}~\bibnamefont{Venugopalan}},
  \bibinfo{journal}{Phys. Rev. Lett.} \textbf{\bibinfo{volume}{84}},
  \bibinfo{pages}{4309} (\bibinfo{year}{2000}), \eprint{hep-ph/9909203}.

\bibitem[{\citenamefont{Krasnitz and Venugopalan}(2001)}]{Krasnitz:2000gz}
\bibinfo{author}{\bibfnamefont{A.}~\bibnamefont{Krasnitz}} \bibnamefont{and}
  \bibinfo{author}{\bibfnamefont{R.}~\bibnamefont{Venugopalan}},
  \bibinfo{journal}{Phys. Rev. Lett.} \textbf{\bibinfo{volume}{86}},
  \bibinfo{pages}{1717} (\bibinfo{year}{2001}), \eprint{hep-ph/0007108}.

\bibitem[{\citenamefont{Krasnitz et~al.}(2001)\citenamefont{Krasnitz, Nara, and
  Venugopalan}}]{Krasnitz:2001qu}
\bibinfo{author}{\bibfnamefont{A.}~\bibnamefont{Krasnitz}},
  \bibinfo{author}{\bibfnamefont{Y.}~\bibnamefont{Nara}}, \bibnamefont{and}
  \bibinfo{author}{\bibfnamefont{R.}~\bibnamefont{Venugopalan}},
  \bibinfo{journal}{Phys. Rev. Lett.} \textbf{\bibinfo{volume}{87}},
  \bibinfo{pages}{192302} (\bibinfo{year}{2001}), \eprint{hep-ph/0108092}.

\bibitem[{\citenamefont{Krasnitz
  et~al.}(2003{\natexlab{a}})\citenamefont{Krasnitz, Nara, and
  Venugopalan}}]{Krasnitz:2002mn}
\bibinfo{author}{\bibfnamefont{A.}~\bibnamefont{Krasnitz}},
  \bibinfo{author}{\bibfnamefont{Y.}~\bibnamefont{Nara}}, \bibnamefont{and}
  \bibinfo{author}{\bibfnamefont{R.}~\bibnamefont{Venugopalan}},
  \bibinfo{journal}{Nucl. Phys.} \textbf{\bibinfo{volume}{A717}},
  \bibinfo{pages}{268} (\bibinfo{year}{2003}{\natexlab{a}}),
  \eprint{hep-ph/0209269}.

\bibitem[{\citenamefont{Krasnitz
  et~al.}(2003{\natexlab{b}})\citenamefont{Krasnitz, Nara, and
  Venugopalan}}]{Krasnitz:2003jw}
\bibinfo{author}{\bibfnamefont{A.}~\bibnamefont{Krasnitz}},
  \bibinfo{author}{\bibfnamefont{Y.}~\bibnamefont{Nara}}, \bibnamefont{and}
  \bibinfo{author}{\bibfnamefont{R.}~\bibnamefont{Venugopalan}},
  \bibinfo{journal}{Nucl. Phys.} \textbf{\bibinfo{volume}{A727}},
  \bibinfo{pages}{427} (\bibinfo{year}{2003}{\natexlab{b}}),
  \eprint{hep-ph/0305112}.

\bibitem[{\citenamefont{Krasnitz
  et~al.}(2003{\natexlab{c}})\citenamefont{Krasnitz, Nara, and
  Venugopalan}}]{Krasnitz:2002ng}
\bibinfo{author}{\bibfnamefont{A.}~\bibnamefont{Krasnitz}},
  \bibinfo{author}{\bibfnamefont{Y.}~\bibnamefont{Nara}}, \bibnamefont{and}
  \bibinfo{author}{\bibfnamefont{R.}~\bibnamefont{Venugopalan}},
  \bibinfo{journal}{Phys. Lett.} \textbf{\bibinfo{volume}{B554}},
  \bibinfo{pages}{21} (\bibinfo{year}{2003}{\natexlab{c}}),
  \eprint{hep-ph/0204361}.

\bibitem[{\citenamefont{Lappi}(2003)}]{Lappi:2003bi}
\bibinfo{author}{\bibfnamefont{T.}~\bibnamefont{Lappi}},
  \bibinfo{journal}{Phys. Rev.} \textbf{\bibinfo{volume}{C67}},
  \bibinfo{pages}{054903} (\bibinfo{year}{2003}), \eprint{hep-ph/0303076}.

\bibitem[{\citenamefont{Lappi and McLerran}(2006)}]{Lappi:2006fp}
\bibinfo{author}{\bibfnamefont{T.}~\bibnamefont{Lappi}} \bibnamefont{and}
  \bibinfo{author}{\bibfnamefont{L.}~\bibnamefont{McLerran}}
  (\bibinfo{year}{2006}), \eprint{hep-ph/0602189}.

\bibitem[{\citenamefont{Fries et~al.}(2006)\citenamefont{Fries, Kapusta, and
  Li}}]{Fries:2006pv}
\bibinfo{author}{\bibfnamefont{R.~J.} \bibnamefont{Fries}},
  \bibinfo{author}{\bibfnamefont{J.~I.} \bibnamefont{Kapusta}},
  \bibnamefont{and} \bibinfo{author}{\bibfnamefont{Y.}~\bibnamefont{Li}}
  (\bibinfo{year}{2006}), \eprint{nucl-th/0604054}.

\bibitem[{\citenamefont{Arnold and Moore}(2006{\natexlab{a}})}]{Arnold:2005ef}
\bibinfo{author}{\bibfnamefont{P.}~\bibnamefont{Arnold}} \bibnamefont{and}
  \bibinfo{author}{\bibfnamefont{G.~D.} \bibnamefont{Moore}},
  \bibinfo{journal}{Phys. Rev.} \textbf{\bibinfo{volume}{D73}},
  \bibinfo{pages}{025006} (\bibinfo{year}{2006}{\natexlab{a}}),
  \eprint{hep-ph/0509206}.

\bibitem[{\citenamefont{Arnold and Moore}(2006{\natexlab{b}})}]{Arnold:2005qs}
\bibinfo{author}{\bibfnamefont{P.}~\bibnamefont{Arnold}} \bibnamefont{and}
  \bibinfo{author}{\bibfnamefont{G.~D.} \bibnamefont{Moore}},
  \bibinfo{journal}{Phys. Rev.} \textbf{\bibinfo{volume}{D73}},
  \bibinfo{pages}{025013} (\bibinfo{year}{2006}{\natexlab{b}}),
  \eprint{hep-ph/0509226}.

\bibitem[{\citenamefont{Kelly et~al.}(1994{\natexlab{a}})\citenamefont{Kelly,
  Liu, Lucchesi, and Manuel}}]{Kelly:1994ig}
\bibinfo{author}{\bibfnamefont{P.~F.} \bibnamefont{Kelly}},
  \bibinfo{author}{\bibfnamefont{Q.}~\bibnamefont{Liu}},
  \bibinfo{author}{\bibfnamefont{C.}~\bibnamefont{Lucchesi}}, \bibnamefont{and}
  \bibinfo{author}{\bibfnamefont{C.}~\bibnamefont{Manuel}},
  \bibinfo{journal}{Phys. Rev. Lett.} \textbf{\bibinfo{volume}{72}},
  \bibinfo{pages}{3461} (\bibinfo{year}{1994}{\natexlab{a}}),
  \eprint{hep-ph/9403403}.

\bibitem[{\citenamefont{Kelly et~al.}(1994{\natexlab{b}})\citenamefont{Kelly,
  Liu, Lucchesi, and Manuel}}]{Kelly:1994dh}
\bibinfo{author}{\bibfnamefont{P.~F.} \bibnamefont{Kelly}},
  \bibinfo{author}{\bibfnamefont{Q.}~\bibnamefont{Liu}},
  \bibinfo{author}{\bibfnamefont{C.}~\bibnamefont{Lucchesi}}, \bibnamefont{and}
  \bibinfo{author}{\bibfnamefont{C.}~\bibnamefont{Manuel}},
  \bibinfo{journal}{Phys. Rev.} \textbf{\bibinfo{volume}{D50}},
  \bibinfo{pages}{4209} (\bibinfo{year}{1994}{\natexlab{b}}),
  \eprint{hep-ph/9406285}.

\bibitem[{\citenamefont{Blaizot and Iancu}(1999)}]{Blaizot:1999xk}
\bibinfo{author}{\bibfnamefont{J.-P.} \bibnamefont{Blaizot}} \bibnamefont{and}
  \bibinfo{author}{\bibfnamefont{E.}~\bibnamefont{Iancu}},
  \bibinfo{journal}{Nucl. Phys.} \textbf{\bibinfo{volume}{B557}},
  \bibinfo{pages}{183} (\bibinfo{year}{1999}), \eprint{hep-ph/9903389}.

\bibitem[{\citenamefont{Matinyan et~al.}(1988)\citenamefont{Matinyan,
  Prokhorenko, and Savvidy}}]{Matinyan:1986hc}
\bibinfo{author}{\bibfnamefont{S.~G.} \bibnamefont{Matinyan}},
  \bibinfo{author}{\bibfnamefont{E.~B.} \bibnamefont{Prokhorenko}},
  \bibnamefont{and} \bibinfo{author}{\bibfnamefont{G.~K.}
  \bibnamefont{Savvidy}}, \bibinfo{journal}{Nucl. Phys.}
  \textbf{\bibinfo{volume}{B298}}, \bibinfo{pages}{414} (\bibinfo{year}{1988}).

\bibitem[{\citenamefont{Kawabe and Ohta}(1990)}]{Kawabe:1988jc}
\bibinfo{author}{\bibfnamefont{T.}~\bibnamefont{Kawabe}} \bibnamefont{and}
  \bibinfo{author}{\bibfnamefont{S.}~\bibnamefont{Ohta}},
  \bibinfo{journal}{Phys. Rev.} \textbf{\bibinfo{volume}{D41}},
  \bibinfo{pages}{1983} (\bibinfo{year}{1990}).

\bibitem[{\citenamefont{Biro et~al.}(1994)\citenamefont{Biro, Gong, M{\"u}ller,
  and Trayanov}}]{Biro:1993qc}
\bibinfo{author}{\bibfnamefont{T.~S.} \bibnamefont{Biro}},
  \bibinfo{author}{\bibfnamefont{C.}~\bibnamefont{Gong}},
  \bibinfo{author}{\bibfnamefont{B.}~\bibnamefont{M{\"u}ller}},
  \bibnamefont{and} \bibinfo{author}{\bibfnamefont{A.}~\bibnamefont{Trayanov}},
  \bibinfo{journal}{Int. J. Mod. Phys.} \textbf{\bibinfo{volume}{C5}},
  \bibinfo{pages}{113} (\bibinfo{year}{1994}), \eprint{nucl-th/9306002}.

\bibitem[{\citenamefont{Wong}(1970)}]{Wong:1970fu}
\bibinfo{author}{\bibfnamefont{S.~K.} \bibnamefont{Wong}},
  \bibinfo{journal}{Nuovo Cim.} \textbf{\bibinfo{volume}{A65S10}},
  \bibinfo{pages}{689} (\bibinfo{year}{1970}).

\bibitem[{\citenamefont{Heinz}(1983)}]{Heinz:1983nx}
\bibinfo{author}{\bibfnamefont{U.~W.} \bibnamefont{Heinz}},
  \bibinfo{journal}{Phys. Rev. Lett.} \textbf{\bibinfo{volume}{51}},
  \bibinfo{pages}{351} (\bibinfo{year}{1983}).

\bibitem[{\citenamefont{Laine and Manuel}(2002)}]{Laine:2001my}
\bibinfo{author}{\bibfnamefont{M.}~\bibnamefont{Laine}} \bibnamefont{and}
  \bibinfo{author}{\bibfnamefont{C.}~\bibnamefont{Manuel}},
  \bibinfo{journal}{Phys. Rev.} \textbf{\bibinfo{volume}{D65}},
  \bibinfo{pages}{077902} (\bibinfo{year}{2002}), \eprint{hep-ph/0111113}.

\bibitem[{\citenamefont{Ambj{\o}rn et~al.}(1991)\citenamefont{Ambj{\o}rn,
  Askgaard, Porter, and Shaposhnikov}}]{Ambjorn:1990pu}
\bibinfo{author}{\bibfnamefont{J.}~\bibnamefont{Ambj{\o}rn}},
  \bibinfo{author}{\bibfnamefont{T.}~\bibnamefont{Askgaard}},
  \bibinfo{author}{\bibfnamefont{H.}~\bibnamefont{Porter}}, \bibnamefont{and}
  \bibinfo{author}{\bibfnamefont{M.~E.} \bibnamefont{Shaposhnikov}},
  \bibinfo{journal}{Nucl. Phys.} \textbf{\bibinfo{volume}{B353}},
  \bibinfo{pages}{346} (\bibinfo{year}{1991}).

\bibitem[{\citenamefont{Hu and M{\"u}ller}(1997)}]{Hu:1996sf}
\bibinfo{author}{\bibfnamefont{C.~R.} \bibnamefont{Hu}} \bibnamefont{and}
  \bibinfo{author}{\bibfnamefont{B.}~\bibnamefont{M{\"u}ller}},
  \bibinfo{journal}{Phys. Lett.} \textbf{\bibinfo{volume}{B409}},
  \bibinfo{pages}{377} (\bibinfo{year}{1997}), \eprint{hep-ph/9611292}.

\bibitem[{\citenamefont{Moore et~al.}(1998)\citenamefont{Moore, Hu, and
  M{\"u}ller}}]{Moore:1997sn}
\bibinfo{author}{\bibfnamefont{G.~D.} \bibnamefont{Moore}},
  \bibinfo{author}{\bibfnamefont{C.-r.} \bibnamefont{Hu}}, \bibnamefont{and}
  \bibinfo{author}{\bibfnamefont{B.}~\bibnamefont{M{\"u}ller}},
  \bibinfo{journal}{Phys. Rev.} \textbf{\bibinfo{volume}{D58}},
  \bibinfo{pages}{045001} (\bibinfo{year}{1998}), \eprint{hep-ph/9710436}.

\bibitem[{\citenamefont{Dumitru and Nara}(2005)}]{Dumitru:2005gp}
\bibinfo{author}{\bibfnamefont{A.}~\bibnamefont{Dumitru}} \bibnamefont{and}
  \bibinfo{author}{\bibfnamefont{Y.}~\bibnamefont{Nara}},
  \bibinfo{journal}{Phys. Lett.} \textbf{\bibinfo{volume}{B621}},
  \bibinfo{pages}{89} (\bibinfo{year}{2005}), \eprint{hep-ph/0503121}.

\bibitem[{\citenamefont{Dumitru and Nara}(2006)}]{Dumitru:2005hj}
\bibinfo{author}{\bibfnamefont{A.}~\bibnamefont{Dumitru}} \bibnamefont{and}
  \bibinfo{author}{\bibfnamefont{Y.}~\bibnamefont{Nara}},
  \bibinfo{journal}{Eur. Phys. J.} \textbf{\bibinfo{volume}{A29}},
  \bibinfo{pages}{65} (\bibinfo{year}{2006}), \eprint{hep-ph/0511242}.

\bibitem[{\citenamefont{Nara}(2006)}]{Nara:2005fr}
\bibinfo{author}{\bibfnamefont{Y.}~\bibnamefont{Nara}}, \bibinfo{journal}{Nucl.
  Phys.} \textbf{\bibinfo{volume}{A774}}, \bibinfo{pages}{783}
  (\bibinfo{year}{2006}), \eprint{nucl-th/0509052}.

\bibitem[{\citenamefont{Hockney and Eastwood}(1981)}]{Hockney:1985}
\bibinfo{author}{\bibfnamefont{R.~W.} \bibnamefont{Hockney}} \bibnamefont{and}
  \bibinfo{author}{\bibfnamefont{J.~W.} \bibnamefont{Eastwood}},
  \emph{\bibinfo{title}{Computer Simulation Using Particles}}
  (\bibinfo{publisher}{McGraw-Hill, New York}, \bibinfo{year}{1981}).

\bibitem[{\citenamefont{Birdsall and Langdon}(1985)}]{Birdsall:1985}
\bibinfo{author}{\bibfnamefont{C.~K.} \bibnamefont{Birdsall}} \bibnamefont{and}
  \bibinfo{author}{\bibfnamefont{A.}~\bibnamefont{Langdon}},
  \emph{\bibinfo{title}{Plasma Physics via Computer Simulation}}
  (\bibinfo{publisher}{McGraw-Hill, New York}, \bibinfo{year}{1985}).

\bibitem[{\citenamefont{Eastwood}(1991)}]{Eastwood:1991}
\bibinfo{author}{\bibfnamefont{J.~W.} \bibnamefont{Eastwood}},
  \bibinfo{journal}{Comput. Phys. Comm.} \textbf{\bibinfo{volume}{64}},
  \bibinfo{pages}{252} (\bibinfo{year}{1991}).

\bibitem[{\citenamefont{Eastwood et~al.}(1995)\citenamefont{Eastwood, Arter,
  Brealey, and Hockney}}]{Eastwood:1995}
\bibinfo{author}{\bibfnamefont{J.~W.} \bibnamefont{Eastwood}},
  \bibinfo{author}{\bibfnamefont{W.}~\bibnamefont{Arter}},
  \bibinfo{author}{\bibfnamefont{N.~J.} \bibnamefont{Brealey}},
  \bibnamefont{and} \bibinfo{author}{\bibfnamefont{R.}~\bibnamefont{Hockney}},
  \bibinfo{journal}{Comput. Phys. Comm.} \textbf{\bibinfo{volume}{87}},
  \bibinfo{pages}{155} (\bibinfo{year}{1995}).

\bibitem[{\citenamefont{Buneman and Villasenor}(1992)}]{Buneman:1992}
\bibinfo{author}{\bibfnamefont{O.}~\bibnamefont{Buneman}} \bibnamefont{and}
  \bibinfo{author}{\bibfnamefont{J.}~\bibnamefont{Villasenor}},
  \bibinfo{journal}{Comput. Phys. Comm.} \textbf{\bibinfo{volume}{69}},
  \bibinfo{pages}{306} (\bibinfo{year}{1992}).

\bibitem[{\citenamefont{Esirkepov}(2001)}]{Esirkepov:2001}
\bibinfo{author}{\bibfnamefont{T.~Z.} \bibnamefont{Esirkepov}},
  \bibinfo{journal}{Comput. Phys. Comm.} \textbf{\bibinfo{volume}{135}},
  \bibinfo{pages}{144} (\bibinfo{year}{2001}).

\bibitem[{\citenamefont{Umeda et~al.}(2003)\citenamefont{Umeda, Omura,
  Tominaga, and Matsumoto}}]{Umeda:2003}
\bibinfo{author}{\bibfnamefont{T.}~\bibnamefont{Umeda}},
  \bibinfo{author}{\bibfnamefont{Y.}~\bibnamefont{Omura}},
  \bibinfo{author}{\bibfnamefont{T.}~\bibnamefont{Tominaga}}, \bibnamefont{and}
  \bibinfo{author}{\bibfnamefont{H.}~\bibnamefont{Matsumoto}},
  \bibinfo{journal}{Comput. Phys. Comm.} \textbf{\bibinfo{volume}{156}},
  \bibinfo{pages}{73} (\bibinfo{year}{2003}).

\bibitem[{\citenamefont{Mueller et~al.}(2006)\citenamefont{Mueller, Shoshi, and
  Wong}}]{Mueller:2006up}
\bibinfo{author}{\bibfnamefont{A.~H.} \bibnamefont{Mueller}},
  \bibinfo{author}{\bibfnamefont{A.~I.} \bibnamefont{Shoshi}},
  \bibnamefont{and} \bibinfo{author}{\bibfnamefont{S.~M.~H.}
  \bibnamefont{Wong}} (\bibinfo{year}{2006}), \eprint{hep-ph/0607136}.

\bibitem[{\citenamefont{B{\"o}deker}(2005)}]{Bodeker:2005nv}
\bibinfo{author}{\bibfnamefont{D.}~\bibnamefont{B{\"o}deker}},
  \bibinfo{journal}{JHEP} \textbf{\bibinfo{volume}{10}}, \bibinfo{pages}{092}
  (\bibinfo{year}{2005}), \eprint{hep-ph/0508223}.

\bibitem[{\citenamefont{Schenke et~al.}(2006)\citenamefont{Schenke, Strickland,
  Greiner, and Thoma}}]{Schenke:2006xu}
\bibinfo{author}{\bibfnamefont{B.}~\bibnamefont{Schenke}},
  \bibinfo{author}{\bibfnamefont{M.}~\bibnamefont{Strickland}},
  \bibinfo{author}{\bibfnamefont{C.}~\bibnamefont{Greiner}}, \bibnamefont{and}
  \bibinfo{author}{\bibfnamefont{M.~H.} \bibnamefont{Thoma}},
  \bibinfo{journal}{Phys. Rev. D} \textbf{\bibinfo{volume}{73}},
  \bibinfo{eid}{125004} (\bibinfo{year}{2006}), \eprint{hep-ph/0603029}.

\bibitem[{\citenamefont{Mueller et~al.}(2005)\citenamefont{Mueller, Shoshi, and
  Wong}}]{Mueller:2005hj}
\bibinfo{author}{\bibfnamefont{A.~H.} \bibnamefont{Mueller}},
  \bibinfo{author}{\bibfnamefont{A.~I.} \bibnamefont{Shoshi}},
  \bibnamefont{and} \bibinfo{author}{\bibfnamefont{S.~M.~H.}
  \bibnamefont{Wong}} (\bibinfo{year}{2005}), \eprint{hep-ph/0512045}.

\end{thebibliography}

\end{document}